\input harvmac
\input amssym
\input epsf

\def\unit{\relax{\rm 1\kern-.26em I}}
\def\nada{\relax{\rm 0\kern-.30em l}}
\def\tilde{\widetilde}

\def\unit{\relax{\rm 1\kern-.26em I}}
\def\nada{\relax{\rm 0\kern-.30em l}}
\def\tilde{\widetilde}

\def\p{\partial}

\def\p{\partial}

\def\unit{\relax{\rm 1\kern-.26em I}}
\def\nada{\relax{\rm 0\kern-.30em l}}
\def\tilde{\widetilde}


\def\p{\partial}

\noblackbox
\def\IL{\relax{\rm I\kern-.18em L}}
\def\IH{\relax{\rm I\kern-.18em H}}
\def\IR{\relax{\rm I\kern-.18em R}}
\def\IC{\relax\hbox{$\inbar\kern-.3em{\rm C}$}}
\def\IZ{\relax\ifmmode\mathchoice
{\hbox{\cmss Z\kern-.4em Z}}{\hbox{\cmss Z\kern-.4em Z}}
{\lower.9pt\hbox{\cmsss Z\kern-.4em Z}} {\lower1.2pt\hbox{\cmsss
Z\kern-.4em Z}}\else{\cmss Z\kern-.4em Z}\fi}

\def\n{\nu}
\def\p{\partial}


\hfill

\def\{{\lbrace}
\def\}{\rbrace}

\def\frac#1#2{{\scriptstyle{#1}\over\scriptstyle{#2}}}

\font\manual=manfnt \def\dbend{\lower3.5pt\hbox{\manual\char127}}

\def\IZ{\relax\ifmmode\mathchoice
{\hbox{\cmss Z\kern-.4em Z}}{\hbox{\cmss Z\kern-.4em Z}}
{\lower.9pt\hbox{\cmsss Z\kern-.4em Z}} {\lower1.2pt\hbox{\cmsss
Z\kern-.4em Z}}\else{\cmss Z\kern-.4em Z}\fi}
\def\half {{1\over 2}}

\def\p{\partial}

\def\rt2{\sqrt{2}}
\def\irt2{{1\over\sqrt{2}}}

\def\hat{\widehat}
\def\epsfcheck{\ifx\epsfbox\UnDeFiNeD
\message{(NO epsf.tex, FIGURES WILL BE IGNORED)}
\gdef\figin##1{\vskip2in}\gdef\figins##1{\hskip.5in}
\else\message{(FIGURES WILL BE INCLUDED)}%
\gdef\figin##1{##1}\gdef\figins##1{##1}\fi}
\def\DefWarn#1{}
\def\figinsert{\goodbreak\midinsert}
\def\ifig#1#2#3{\DefWarn#1\xdef#1{fig.~\the\figno}
\writedef{#1\leftbracket fig.\noexpand~\the\figno}%
\figinsert\figin{\centerline{#3}}\medskip\centerline{\vbox{\baselineskip12pt
\advance\hsize by -1truein\noindent\footnotefont{\bf
Fig.~\the\figno:\ } \it#2}}
\bigskip\endinsert\global\advance\figno by1}


\def\p{\partial}

\noblackbox
\def\IL{\relax{\rm I\kern-.18em L}}
\def\IH{\relax{\rm I\kern-.18em H}}
\def\IR{\relax{\rm I\kern-.18em R}}
\def\IC{\relax\hbox{$\inbar\kern-.3em{\rm C}$}}
\def\IZ{\relax\ifmmode\mathchoice
{\hbox{\cmss Z\kern-.4em Z}}{\hbox{\cmss Z\kern-.4em Z}}
{\lower.9pt\hbox{\cmsss Z\kern-.4em Z}} {\lower1.2pt\hbox{\cmsss
Z\kern-.4em Z}}\else{\cmss Z\kern-.4em Z}\fi}

\def\CF {{\cal F}}

\def\CO {{\cal O}}

\def\CC {{\cal C}}

\def\CA{{\cal A}}


\def\CO {{\cal O}}

\font\manual=manfnt \def\dbend{\lower3.5pt\hbox{\manual\char127}}

\def\IZ{\relax\ifmmode\mathchoice
{\hbox{\cmss Z\kern-.4em Z}}{\hbox{\cmss Z\kern-.4em Z}}
{\lower.9pt\hbox{\cmsss Z\kern-.4em Z}} {\lower1.2pt\hbox{\cmsss
Z\kern-.4em Z}}\else{\cmss Z\kern-.4em Z}\fi}
\def\half {{1\over 2}}

\def\p{\partial}

\def\rt2{\sqrt{2}}
\def\irt2{{1\over\sqrt{2}}}

\def\hat{\widehat}

\def\figin{\epsfcheck\figin}\def\figins{\epsfcheck\figins}
\def\epsfcheck{\ifx\epsfbox\UnDeFiNeD
\message{(NO epsf.tex, FIGURES WILL BE IGNORED)}
\gdef\figin##1{\vskip2in}\gdef\figins##1{\hskip.5in}
\else\message{(FIGURES WILL BE INCLUDED)}%
\gdef\figin##1{##1}\gdef\figins##1{##1}\fi}
\def\DefWarn#1{}
\def\figinsert{\goodbreak\midinsert}
\def\ifig#1#2#3{\DefWarn#1\xdef#1{fig.~\the\figno}
\writedef{#1\leftbracket fig.\noexpand~\the\figno}%
\figinsert\figin{\centerline{#3}}\medskip\centerline{\vbox{\baselineskip12pt
\advance\hsize by -1truein\noindent\footnotefont{\bf
Fig.~\the\figno:\ } \it#2}}
\bigskip\endinsert\global\advance\figno by1}


\lref\CappelliYC{
  A.~Cappelli, D.~Friedan and J.~I.~Latorre,
Nucl.\ Phys.\ B {\bf 352}, 616 (1991)..
}

\lref\IorioAD{
  A.~Iorio, L.~O'Raifeartaigh, I.~Sachs and C.~Wiesendanger,
Nucl.\ Phys.\ B {\bf 495}, 433 (1997).
[hep-th/9607110].
}

\lref\IntriligatorJJ{
  K.~A.~Intriligator, B.~Wecht,
  ``The Exact Superconformal R Symmetry Maximizes a,''
Nucl.\ Phys.\  {\bf B667}, 183-200 (2003). [hep-th/0304128].
}

\lref\book{
  R.~J~Eden, P.~V.~Landshoff, D.~I.~Olive, and J.~C.~Polkinghorne,
  ``The Analytic S-Matrix,''
Cambridge University Press, 1966. 
}

\lref\SeibergPQ{
  N.~Seiberg,
  ``Electric - Magnetic Duality in Supersymmetric nonAbelian Gauge Theories,''
Nucl.\ Phys.\  {\bf B435}, 129-146 (1995). [hep-th/9411149].
}

\lref\KutasovIY{
  D.~Kutasov, A.~Parnachev, D.~A.~Sahakyan,
 ``Central charges and U(1)(R) symmetries in N=1 superYang-Mills,''
JHEP {\bf 0311}, 013 (2003). [hep-th/0308071].
}

\lref\IntriligatorMI{
  K.~A.~Intriligator, B.~Wecht,
  ``RG fixed points and flows in SQCD with adjoints,''
Nucl.\ Phys.\  {\bf B677}, 223-272 (2004). [hep-th/0309201].
}

\lref\WittenTW{
  E.~Witten,
  ``Global Aspects of Current Algebra,''
Nucl.\ Phys.\  {\bf B223}, 422-432 (1983).
}

\lref\FrishmanDQ{
  Y.~Frishman, A.~Schwimmer, T.~Banks, S.~Yankielowicz,
  ``The Axial Anomaly and the Bound State Spectrum in Confining Theories,''
Nucl.\ Phys.\  {\bf B177}, 157 (1981).
}

\lref\AppelquistHR{
  T.~Appelquist, A.~G.~Cohen, M.~Schmaltz,
  ``A New Constraint on Strongly Coupled Gauge Theories,''
Phys.\ Rev.\  {\bf D60}, 045003 (1999). [arXiv:hep-th/9901109
[hep-th]].
}
\lref\AnselmiAM{
  D.~Anselmi, D.~Z.~Freedman, M.~T.~Grisaru, A.~A.~Johansen,
  ``Nonperturbative Formulas for Central Functions of Supersymmetric Gauge Theories,''
Nucl.\ Phys.\  {\bf B526}, 543-571 (1998). [hep-th/9708042].
}

\lref\AnselmiYS{
  D.~Anselmi, J.~Erlich, D.~Z.~Freedman, A.~A.~Johansen,
  ``Positivity Constraints on Anomalies in Supersymmetric Gauge Theories,''
Phys.\ Rev.\  {\bf D57}, 7570-7588 (1998). [hep-th/9711035].
}

\lref\CardyCWA{
  J.~L.~Cardy,
  ``Is There a c Theorem in Four-Dimensions?,''
Phys.\ Lett.\  {\bf B215}, 749-752 (1988). }

\lref\OsbornTD{
  H.~Osborn,
  ``Derivation of a Four-Dimensional c Theorem,''
Phys.\ Lett.\  {\bf B222}, 97 (1989).
}

\lref\KutasovIY{
  D.~Kutasov, A.~Parnachev, D.~A.~Sahakyan,
  ``Central Charges and U(1)(R) Symmetries in N=1 SuperYang-Mills,''
JHEP {\bf 0311}, 013 (2003). [hep-th/0308071].
}

\lref\IntriligatorMI{
  K.~A.~Intriligator, B.~Wecht,
  ``RG Fixed Points and Flows in SQCD with Adjoints,''
Nucl.\ Phys.\  {\bf B677}, 223-272 (2004). [hep-th/0309201].
}

\lref\JackEB{
  I.~Jack, H.~Osborn,
  ``Analogs for the C Theorem for Four-dimensional Renormalizable Field Theories,''
Nucl.\ Phys.\  {\bf B343}, 647-688 (1990).
}

\lref\ZamolodchikovGT{
  A.~B.~Zamolodchikov,
  ``Irreversibility of the Flux of the Renormalization Group in a 2D Field Theory,''
JETP Lett.\  {\bf 43}, 730-732 (1986). }

\lref\WessYU{
  J.~Wess, B.~Zumino,
  ``Consequences of Anomalous Ward Identities,''
Phys.\ Lett.\  {\bf B37}, 95 (1971). }

\lref\SchwimmerZA{
  A.~Schwimmer, S.~Theisen,
  ``Spontaneous Breaking of Conformal Invariance and Trace Anomaly Matching,''
Nucl.\ Phys.\  {\bf B847}, 590-611 (2011). [arXiv:1011.0696
[hep-th]].
}

\lref\PhamCR{
  T.~N.~Pham, T.~N.~Truong,
  ``Evaluation of the Derivative Quartic Terms of The Meson Chiral Lagrangian from Forward Dispersion Relation,''
Phys.\ Rev.\  {\bf D31}, 3027 (1985).
}

\lref\AdamsSV{
  A.~Adams, N.~Arkani-Hamed, S.~Dubovsky, A.~Nicolis, R.~Rattazzi,
  ``Causality, Analyticity and an IR Obstruction to UV Completion,''
JHEP {\bf 0610}, 014 (2006). [hep-th/0602178].
}

\lref\DineSW{
  M.~Dine, G.~Festuccia, Z.~Komargodski,
  ``A Bound on the Superpotential,''
JHEP {\bf 1003}, 011 (2010). [arXiv:0910.2527 [hep-th]].
}

\lref\CappelliYC{
  A.~Cappelli, D.~Friedan, J.~I.~Latorre,
  ``C Theorem and Spectral Representation,''
Nucl.\ Phys.\  {\bf B352}, 616-670 (1991).
}

\lref\FrishmanDQ{
  Y.~Frishman, A.~Schwimmer, T.~Banks, S.~Yankielowicz,
  ``The Axial Anomaly and the Bound State Spectrum in Confining Theories,''
Nucl.\ Phys.\  {\bf B177}, 157 (1981).
}

\lref\WittenTW{
  E.~Witten,
  ``Global Aspects of Current Algebra,''
Nucl.\ Phys.\  {\bf B223}, 422-432 (1983).
}

\lref\BirrellIX{
  N.~D.~Birrell, P.~C.~W.~Davies,
  ``Quantum Fields in Curved Space,''
Cambridge, Uk: Univ. Pr. ( 1982) 340p.
}

\lref\MyersTJ{
  R.~C.~Myers, A.~Sinha,
 ``Holographic c-Theorems in Arbitrary Dimensions,''
JHEP {\bf 1101}, 125 (2011).
[arXiv:1011.5819 [hep-th]].
}

\lref\DuffWM{
  M.~J.~Duff,
  ``Twenty Years of the Weyl Anomaly,''
Class.\ Quant.\ Grav.\  {\bf 11}, 1387-1404 (1994).
[hep-th/9308075].
}

\lref\BuchbinderJN{
  I.~L.~Buchbinder, S.~M.~Kuzenko, A.~A.~Tseytlin,
 ``On Low-Energy Effective Actions in N=2, N=4 Superconformal Theories in Four-Dimensions,''
Phys.\ Rev.\  {\bf D62}, 045001 (2000).
[hep-th/9911221].
}

\lref\JafferisZI{
  D.~L.~Jafferis, I.~R.~Klebanov, S.~S.~Pufu, B.~R.~Safdi,
  ``Towards the F-Theorem: N=2 Field Theories on the Three-Sphere,''
JHEP {\bf 1106}, 102 (2011).
[arXiv:1103.1181 [hep-th]].
}

\lref\RiegertKT{
  R.~J.~Riegert,
 ``A Nonlocal Action for the Trace Anomaly,''
Phys.\ Lett.\  {\bf B134}, 56-60 (1984).
}

\lref\FradkinTG{
  E.~S.~Fradkin, A.~A.~Tseytlin,
 ``Conformal Anomaly in Weyl Theory and Anomaly Free Superconformal Theories,''
Phys.\ Lett.\  {\bf B134}, 187 (1984).
}

\lref\TS{
  A.~Schwimmer, S.~Theisen,
 in preparation.}

\lref\KulaxiziJT{
  M.~Kulaxizi, A.~Parnachev,
  ``Energy Flux Positivity and Unitarity in CFTs,''
Phys.\ Rev.\ Lett.\  {\bf 106}, 011601 (2011).
[arXiv:1007.0553 [hep-th]].
}

\lref\KlebanovGS{
  I.~R.~Klebanov, S.~S.~Pufu, B.~R.~Safdi,
 ``F-Theorem without Supersymmetry,''
[arXiv:1105.4598 [hep-th]].
}

\lref\HofmanAR{
  D.~M.~Hofman, J.~Maldacena,
  ``Conformal Collider Physics: Energy and Charge Correlations,''
JHEP {\bf 0805}, 012 (2008).
[arXiv:0803.1467 [hep-th]].
}

\lref\GrinsteinQK{
  B.~Grinstein, K.~A.~Intriligator and I.~Z.~Rothstein,
  ``Comments on Unparticles,''
Phys.\ Lett.\ B {\bf 662}, 367 (2008).
[arXiv:0801.1140 [hep-ph]].
}

\lref\FradkinYW{
  E.~S.~Fradkin, G.~A.~Vilkovisky,
  ``Conformal Off Mass Shell Extension and Elimination of Conformal Anomalies in Quantum Gravity,''
Phys.\ Lett.\  {\bf B73}, 209-213 (1978).
}

\lref\DeserYX{
  S.~Deser, A.~Schwimmer,
  ``Geometric Classification of Conformal Anomalies in Arbitrary Dimensions,''
Phys.\ Lett.\  {\bf B309}, 279-284 (1993).
[hep-th/9302047].
}

\lref\BonoraCQ{
  L.~Bonora, P.~Pasti, M.~Bregola,
  ``Weyl Cocycles,''
Class.\ Quant.\ Grav.\  {\bf 3}, 635 (1986).
}

\lref\NirSV{
  Y.~Nir,
 ``Infrared Treatment of Higher Anomalies and Their Consequences,''
Phys.\ Rev.\  {\bf D34}, 1164-1168 (1986).
}

\lref\DistlerIF{
  J.~Distler, B.~Grinstein, R.~A.~Porto, I.~Z.~Rothstein,
  ``Falsifying Models of New Physics via WW Scattering,''
Phys.\ Rev.\ Lett.\  {\bf 98}, 041601 (2007). [hep-ph/0604255].
}

\lref\FreedmanGP{
  D.~Z.~Freedman, S.~S.~Gubser, K.~Pilch, N.~P.~Warner,
  ``Renormalization group flows from holography supersymmetry and a c theorem,''
Adv.\ Theor.\ Math.\ Phys.\  {\bf 3}, 363-417 (1999).
[hep-th/9904017].
}

\lref\GirardelloBD{
  L.~Girardello, M.~Petrini, M.~Porrati, A.~Zaffaroni,
  ``The Supergravity dual of N=1 superYang-Mills theory,''
Nucl.\ Phys.\  {\bf B569}, 451-469 (2000). [hep-th/9909047].
}

\lref\GiddingsGJ{
  S.~B.~Giddings and R.~A.~Porto,
  ``The Gravitational S-matrix,''
Phys.\ Rev.\ D {\bf 81}, 025002 (2010).
[arXiv:0908.0004 [hep-th]].
}

\lref\NachtmannMR{
  O.~Nachtmann,
  ``Positivity constraints for anomalous dimensions,''
Nucl.\ Phys.\ B {\bf 63}, 237 (1973)..
}

\lref\HeemskerkPN{
  I.~Heemskerk, J.~Penedones, J.~Polchinski and J.~Sully,
  ``Holography from Conformal Field Theory,''
JHEP {\bf 0910}, 079 (2009).
[arXiv:0907.0151 [hep-th]].
}

\lref\FitzpatrickZM{
  A.~L.~Fitzpatrick, E.~Katz, D.~Poland and D.~Simmons-Duffin,
  ``Effective Conformal Theory and the Flat-Space Limit of AdS,''
JHEP {\bf 1107}, 023 (2011).
[arXiv:1007.2412 [hep-th]].
}

\lref\NachtmannMR{
  O.~Nachtmann,
  ``Positivity constraints for anomalous dimensions,''
Nucl.\ Phys.\ B {\bf 63}, 237 (1973).
}

\lref\BanksBJ{
  T.~Banks and G.~Festuccia,
  ``The Regge Limit for Green Functions in Conformal Field Theory,''
JHEP {\bf 1006}, 105 (2010).
[arXiv:0910.2746 [hep-th]].
}

\lref\Eden{
  R.~J.~Eden, P.~V.~Landshoff, D.~I.~Olive and J.~C.~Polkinghorne,
  ``The analytic S matrix''
}

\lref\PolyakovGS{
  A.~M.~Polyakov,
  ``Nonhamiltonian approach to conformal quantum field theory,''
Zh.\ Eksp.\ Teor.\ Fiz.\  {\bf 66}, 23 (1974).
}

\lref\HofmanAR{
  D.~M.~Hofman and J.~Maldacena,
  ``Conformal collider physics: Energy and charge correlations,''
JHEP {\bf 0805}, 012 (2008).
[arXiv:0803.1467 [hep-th]].
}

\lref\AldayMF{
  L.~F.~Alday and J.~M.~Maldacena,
  ``Comments on operators with large spin,''
JHEP {\bf 0711}, 019 (2007).
[arXiv:0708.0672 [hep-th]].
}

\lref\AldayZY{
  L.~F.~Alday, B.~Eden, G.~P.~Korchemsky, J.~Maldacena and E.~Sokatchev,
  ``From correlation functions to Wilson loops,''
JHEP {\bf 1109}, 123 (2011).
[arXiv:1007.3243 [hep-th]].
}

\lref\DolanDV{
  F.~A.~Dolan and H.~Osborn,
  ``Conformal Partial Waves: Further Mathematical Results,''
[arXiv:1108.6194 [hep-th]].
}

\lref\ElShowkHT{
  S.~El-Showk, M.~F.~Paulos, D.~Poland, S.~Rychkov, D.~Simmons-Duffin and A.~Vichi,
  ``Solving the 3D Ising Model with the Conformal Bootstrap,''
[arXiv:1203.6064 [hep-th]].
}

\lref\WilsonJJ{
  K.~G.~Wilson and J.~B.~Kogut,
  ``The Renormalization group and the epsilon expansion,''
Phys.\ Rept.\  {\bf 12}, 75 (1974)..
}

\lref\LangZW{
  K.~Lang and W.~Ruhl,
  ``The Critical O(N) sigma model at dimensions 2 < d < 4: Fusion coefficients and anomalous dimensions,''
Nucl.\ Phys.\ B {\bf 400}, 597 (1993)..
}

\lref\HoffmannDX{
  L.~Hoffmann, L.~Mesref and W.~Ruhl,
  ``Conformal partial wave analysis of AdS amplitudes for dilaton axion four point functions,''
Nucl.\ Phys.\ B {\bf 608}, 177 (2001).
[hep-th/0012153].
}

\lref\ElShowkHT{
  S.~El-Showk, M.~F.~Paulos, D.~Poland, S.~Rychkov, D.~Simmons-Duffin and A.~Vichi,
  ``Solving the 3D Ising Model with the Conformal Bootstrap,''
[arXiv:1203.6064 [hep-th]].
}

\lref\GrossUN{
 D. Gross, {\it unpublished} 
}

\lref\CallanZE{
  C.~G.~Callan, Jr., S.~R.~Coleman and R.~Jackiw,
  ``A New improved energy - momentum tensor,''
Annals Phys.\  {\bf 59}, 42 (1970).
}

\lref\ColemanJE{
  S.~R.~Cole  man and R.~Jackiw,
  ``Why dilatation generators do not generate dilatations?,''
Annals Phys.\  {\bf 67}, 552 (1971).
}

\lref\CallanPU{
  C.~G.~Callan, Jr. and D.~J.~Gross,
  ``Bjorken scaling in quantum field theory,''
Phys.\ Rev.\ D {\bf 8}, 4383 (1973).
}

\lref\MackJE{
  G.~Mack,
 ``All Unitary Ray Representations of the Conformal Group SU(2,2) with Positive Energy,''
Commun.\ Math.\ Phys.\  {\bf 55}, 1 (1977).
}

\lref\GrinsteinQK{
  B.~Grinstein, K.~A.~Intriligator and I.~Z.~Rothstein,
  ``Comments on Unparticles,''
Phys.\ Lett.\ B {\bf 662}, 367 (2008).
[arXiv:0801.1140 [hep-ph]].
}

\lref\RattazziPE{
  R.~Rattazzi, V.~S.~Rychkov, E.~Tonni and A.~Vichi,
  ``Bounding scalar operator dimensions in 4D CFT,''
JHEP {\bf 0812}, 031 (2008).
[arXiv:0807.0004 [hep-th]].
}

\lref\CostaCB{
  M.~S.~Costa, V.~Goncalves and J.~Penedones,
  ``Conformal Regge theory,''
[arXiv:1209.4355 [hep-th]].
}

\lref\PappadopuloJK{
  D.~Pappadopulo, S.~Rychkov, J.~Espin and R.~Rattazzi,
  ``OPE Convergence in Conformal Field Theory,''
[arXiv:1208.6449 [hep-th]].
}

\lref\AdamsSV{
  A.~Adams, N.~Arkani-Hamed, S.~Dubovsky, A.~Nicolis and R.~Rattazzi,
  ``Causality, analyticity and an IR obstruction to UV completion,''
JHEP {\bf 0610}, 014 (2006).
[hep-th/0602178].
}

\lref\FreyhultMY{
  L.~Freyhult and S.~Zieme,
  ``The virtual scaling function of AdS/CFT,''
Phys.\ Rev.\ D {\bf 79}, 105009 (2009).
[arXiv:0901.2749 [hep-th]].
}

\lref\PelissettoEK{
  A.~Pelissetto and E.~Vicari,
  ``Critical phenomena and renormalization group theory,''
Phys.\ Rept.\  {\bf 368}, 549 (2002).
[cond-mat/0012164].
}

\lref\VasilievBA{
  M.~A.~Vasiliev,
  ``Higher spin gauge theories: Star product and AdS space,''
In *Shifman, M.A. (ed.): The many faces of the superworld* 533-610.
[hep-th/9910096].
}

\lref\MaldacenaRE{
  J.~M.~Maldacena,
  ``The Large N limit of superconformal field theories and supergravity,''
Adv.\ Theor.\ Math.\ Phys.\  {\bf 2}, 231 (1998).
[hep-th/9711200].
}

\lref\GubserBC{
  S.~S.~Gubser, I.~R.~Klebanov and A.~M.~Polyakov,
  ``Gauge theory correlators from noncritical string theory,''
Phys.\ Lett.\ B {\bf 428}, 105 (1998).
[hep-th/9802109].
}

\lref\WittenQJ{
  E.~Witten,
  ``Anti-de Sitter space and holography,''
Adv.\ Theor.\ Math.\ Phys.\  {\bf 2}, 253 (1998).
[hep-th/9802150].
}

\lref\SundborgWP{
  B.~Sundborg,
  ``Stringy gravity, interacting tensionless strings and massless higher spins,''
Nucl.\ Phys.\ Proc.\ Suppl.\  {\bf 102}, 113 (2001).
[hep-th/0103247].
}

\lref\DolanUT{
  F.~A.~Dolan and H.~Osborn,
  ``Conformal four point functions and the operator product expansion,''
Nucl.\ Phys.\ B {\bf 599}, 459 (2001).
[hep-th/0011040].
}

\lref\CornalbaXK{
  L.~Cornalba, M.~S.~Costa, J.~Penedones and R.~Schiappa,
  ``Eikonal Approximation in AdS/CFT: From Shock Waves to Four-Point Functions,''
JHEP {\bf 0708}, 019 (2007).
[hep-th/0611122].
}

\lref\CornalbaXM{
  L.~Cornalba, M.~S.~Costa, J.~Penedones and R.~Schiappa,
  ``Eikonal Approximation in AdS/CFT: Conformal Partial Waves and Finite N Four-Point Functions,''
Nucl.\ Phys.\ B {\bf 767}, 327 (2007).
[hep-th/0611123].
}

\lref\CornalbaZB{
  L.~Cornalba, M.~S.~Costa and J.~Penedones,
  ``Eikonal approximation in AdS/CFT: Resumming the gravitational loop expansion,''
JHEP {\bf 0709}, 037 (2007).
[arXiv:0707.0120 [hep-th]].
}

\lref\DerkachovPH{
  S.~E.~Derkachov and A.~N.~Manashov,
  ``Generic scaling relation in the scalar phi**4 model,''
J.\ Phys.\ A {\bf 29}, 8011 (1996).
[hep-th/9604173].
}

\lref\KlebanovJA{
  I.~R.~Klebanov and A.~M.~Polyakov,
  ``AdS dual of the critical O(N) vector model,''
Phys.\ Lett.\ B {\bf 550}, 213 (2002).
[hep-th/0210114].
}

\lref\SezginRT{
  E.~Sezgin and P.~Sundell,
  ``Massless higher spins and holography,''
Nucl.\ Phys.\ B {\bf 644}, 303 (2002), [Erratum-ibid.\ B {\bf 660}, 403 (2003)].
[hep-th/0205131].
}
\lref\EpsteinBG{
  H.~Epstein, V.~Glaser and A.~Martin,
  ``Polynomial behaviour of scattering amplitudes at fixed momentum transfer in theories with local observables,''
Commun.\ Math.\ Phys.\  {\bf 13}, 257 (1969)..
}

\lref\FortinCQ{
  J.~-F.~Fortin, B.~Grinstein and A.~Stergiou,
  ``Scale without Conformal Invariance in Four Dimensions,''
[arXiv:1206.2921 [hep-th]].
}

\lref\FortinHN{
  J.~-F.~Fortin, B.~Grinstein and A.~Stergiou,
  ``A generalized c-theorem and the consistency of scale without conformal invariance,''
[arXiv:1208.3674 [hep-th]].
}

\lref\FortinHC{
  J.~-F.~Fortin, B.~Grinstein, C.~W.~Murphy and A.~Stergiou,
  ``On Limit Cycles in Supersymmetric Theories,''
[arXiv:1210.2718 [hep-th]].
}

\lref\NakayamaED{
  Y.~Nakayama,
  ``Is boundary conformal in CFT?,''
[arXiv:1210.6439 [hep-th]].
}

\lref\NakayamaND{
  Y.~Nakayama,
  ``Supercurrent, Supervirial and Superimprovement,''
[arXiv:1208.4726 [hep-th]].
}

\lref\AntoniadisGN{
  I.~Antoniadis and M.~Buican,
  ``On R-symmetric Fixed Points and Superconformality,''
Phys.\ Rev.\ D {\bf 83}, 105011 (2011).
[arXiv:1102.2294 [hep-th]].
}

\lref\LutyWW{
  M.~A.~Luty, J.~Polchinski and R.~Rattazzi,
  ``The $a$-theorem and the Asymptotics of 4D Quantum Field Theory,''
[arXiv:1204.5221 [hep-th]].
}

\lref\GiombiWH{
  S.~Giombi and X.~Yin,
  ``Higher Spin Gauge Theory and Holography: The Three-Point Functions,''
JHEP {\bf 1009}, 115 (2010).
[arXiv:0912.3462 [hep-th]].
}

\lref\KomargodskiVJ{
  Z.~Komargodski and A.~Schwimmer,
  ``On Renormalization Group Flows in Four Dimensions,''
JHEP {\bf 1112}, 099 (2011).
[arXiv:1107.3987 [hep-th]].
}

\lref\KomargodskiXV{
  Z.~Komargodski,
  ``The Constraints of Conformal Symmetry on RG Flows,''
JHEP {\bf 1207}, 069 (2012).
[arXiv:1112.4538 [hep-th]].
}

\lref\KehreinIA{
  S.~K.~Kehrein,
  ``The Spectrum of critical exponents in phi**2 in two-dimensions theory in D = (4-epsilon)-dimensions: Resolution of degeneracies and hierarchical structures,''
Nucl.\ Phys.\ B {\bf 453}, 777 (1995).
[hep-th/9507044].
}

\lref\BraunRP{
  V.~M.~Braun, G.~P.~Korchemsky and D.~Mueller,
  ``The Uses of conformal symmetry in QCD,''
Prog.\ Part.\ Nucl.\ Phys.\  {\bf 51}, 311 (2003).
[hep-ph/0306057].
}

\lref\HeemskerkPN{
  I.~Heemskerk, J.~Penedones, J.~Polchinski and J.~Sully,
  ``Holography from Conformal Field Theory,''
JHEP {\bf 0910}, 079 (2009).
[arXiv:0907.0151 [hep-th]].
}

\lref\RuhlBW{
  W.~Ruhl,
  ``The Goldstone fields of interacting higher spin field theory on AdS(4),''
[hep-th/0607197].
}

\lref\AdamsSV{
  A.~Adams, N.~Arkani-Hamed, S.~Dubovsky, A.~Nicolis and R.~Rattazzi,
  ``Causality, analyticity and an IR obstruction to UV completion,''
JHEP {\bf 0610}, 014 (2006).
[hep-th/0602178].
}

\lref\PhamCR{
  T.~N.~Pham and T.~N.~Truong,
  ``Evaluation Of The Derivative Quartic Terms Of The Meson Chiral Lagrangian From Forward Dispersion Relation,''
Phys.\ Rev.\ D {\bf 31}, 3027 (1985)..
}

\lref\FGG{
  S.~Ferrara, A.~F.~Grillo and R.~Gatto,
  ``Tensor representations of conformal algebra and conformally covariant operator product expansion,''
Annals Phys.\  {\bf 76}, 161 (1973).
}

\lref\BPZ{
  A.~A.~Belavin, A.~M.~Polyakov and A.~B.~Zamolodchikov,
  ``Infinite Conformal Symmetry in Two-Dimensional Quantum Field Theory,''
Nucl.\ Phys.\ B {\bf 241}, 333 (1984).
}

\lref\HeslopDU{
  P.~J.~Heslop,
  ``Aspects of superconformal field theories in six dimensions,''
JHEP {\bf 0407}, 056 (2004).
[hep-th/0405245].
}

\lref\DolanTT{
  F.~A.~Dolan and H.~Osborn,
  ``Superconformal symmetry, correlation functions and the operator product expansion,''
Nucl.\ Phys.\ B {\bf 629}, 3 (2002).
[hep-th/0112251].
}

\lref\PappadopuloJK{
  D.~Pappadopulo, S.~Rychkov, J.~Espin and R.~Rattazzi,
  ``OPE Convergence in Conformal Field Theory,''
[arXiv:1208.6449 [hep-th]].
}

\lref\FerraraEJ{
  S.~Ferrara, C.~Fronsdal and A.~Zaffaroni,
  ``On N=8 supergravity on AdS(5) and N=4 superconformal Yang-Mills theory,''
Nucl.\ Phys.\ B {\bf 532}, 153 (1998).
[hep-th/9802203].
}

\lref\DidenkoTV{
  V.~E.~Didenko and E.~D.~Skvortsov,
  ``Exact higher-spin symmetry in CFT: all correlators in unbroken Vasiliev theory,''
[arXiv:1210.7963 [hep-th]].
}

\lref\MaldacenaSF{
  J.~Maldacena and A.~Zhiboedov,
  ``Constraining conformal field theories with a slightly broken higher spin symmetry,''
[arXiv:1204.3882 [hep-th]].
}

\lref\MaldacenaJN{
  J.~Maldacena and A.~Zhiboedov,
  ``Constraining Conformal Field Theories with A Higher Spin Symmetry,''
[arXiv:1112.1016 [hep-th]].
}

\lref\BriganteGZ{
  M.~Brigante, H.~Liu, R.~C.~Myers, S.~Shenker and S.~Yaida,
  ``The Viscosity Bound and Causality Violation,''
Phys.\ Rev.\ Lett.\  {\bf 100}, 191601 (2008).
[arXiv:0802.3318 [hep-th]].
}

\lref\GirardelloPP{
  L.~Girardello, M.~Porrati and A.~Zaffaroni,
  ``3-D interacting CFTs and generalized Higgs phenomenon in higher spin theories on AdS,''
Phys.\ Lett.\ B {\bf 561}, 289 (2003).
[hep-th/0212181].
}

\lref\HeemskerkPN{
  I.~Heemskerk, J.~Penedones, J.~Polchinski and J.~Sully,
  ``Holography from Conformal Field Theory,''
JHEP {\bf 0910}, 079 (2009).
[arXiv:0907.0151 [hep-th]].
}

\lref\CornalbaZB{
  L.~Cornalba, M.~S.~Costa and J.~Penedones,
  ``Eikonal approximation in AdS/CFT: Resumming the gravitational loop expansion,''
JHEP {\bf 0709}, 037 (2007).
[arXiv:0707.0120 [hep-th]].
}

\lref\DolanDV{
  F.~A.~Dolan and H.~Osborn,
  ``Conformal Partial Waves: Further Mathematical Results,''
[arXiv:1108.6194 [hep-th]].
}

\lref\DolanUT{
  F.~A.~Dolan and H.~Osborn,
  ``Conformal four point functions and the operator product expansion,''
Nucl.\ Phys.\ B {\bf 599}, 459 (2001).
[hep-th/0011040].
}

\lref\GiombiKC{
  S.~Giombi, S.~Minwalla, S.~Prakash, S.~P.~Trivedi, S.~R.~Wadia and X.~Yin,
  ``Chern-Simons Theory with Vector Fermion Matter,''
Eur.\ Phys.\ J.\ C {\bf 72}, 2112 (2012).
[arXiv:1110.4386 [hep-th]].
}

\lref\AharonyJZ{
  O.~Aharony, G.~Gur-Ari and R.~Yacoby,
  ``d=3 Bosonic Vector Models Coupled to Chern-Simons Gauge Theories,''
JHEP {\bf 1203}, 037 (2012).
[arXiv:1110.4382 [hep-th]].
}

\lref\KovtunKW{
  P.~Kovtun and A.~Ritz,
  ``Black holes and universality classes of critical points,''
Phys.\ Rev.\ Lett.\  {\bf 100}, 171606 (2008).
[arXiv:0801.2785 [hep-th]].
}

\lref\DorigoniRA{
  D.~Dorigoni and V.~S.~Rychkov,
[arXiv:0910.1087 [hep-th]].
}

\lref\FroissartUX{
  M.~Froissart,
  ``Asymptotic behavior and subtractions in the Mandelstam representation,''
Phys.\ Rev.\  {\bf 123}, 1053 (1961)..
}

\lref\MartinRT{
  A.~Martin,
  ``Unitarity and high-energy behavior of scattering amplitudes,''
Phys.\ Rev.\  {\bf 129}, 1432 (1963)..
}

\lref\FrankfurtMC{
  L.~Frankfurt, M.~Strikman and C.~Weiss,
 ``Small-x physics: From HERA to LHC and beyond,''
Ann.\ Rev.\ Nucl.\ Part.\ Sci.\  {\bf 55}, 403 (2005).
[hep-ph/0507286].
}

\lref\OsbornGM{
  H.~Osborn,
  ``Weyl consistency conditions and a local renormalization group equation for general renormalizable field theories,''
Nucl.\ Phys.\ B {\bf 363}, 486 (1991)..
}

\lref\PolchinskiJW{
  J.~Polchinski and M.~J.~Strassler,
  ``Deep inelastic scattering and gauge / string duality,''
JHEP {\bf 0305}, 012 (2003).
[hep-th/0209211].
}

\lref\BarnesBM{
  E.~Barnes, E.~Gorbatov, K.~A.~Intriligator, M.~Sudano and J.~Wright,
  ``The Exact superconformal R-symmetry minimizes tau(RR),''
Nucl.\ Phys.\ B {\bf 730}, 210 (2005).
[hep-th/0507137].
}

\lref\PolandWG{
  D.~Poland and D.~Simmons-Duffin,
  ``Bounds on 4D Conformal and Superconformal Field Theories,''
JHEP {\bf 1105}, 017 (2011).
[arXiv:1009.2087 [hep-th]].
}

\lref\LiendoHY{
  P.~Liendo, L.~Rastelli and B.~C.~van Rees,
  ``The Bootstrap Program for Boundary $CFT_d$,''
[arXiv:1210.4258 [hep-th]].
}

\lref\PolyakovXD{
 A.~M.~Polyakov,
 ``Conformal symmetry of critical fluctuations,''
JETP Lett.\  {\bf 12}, 381 (1970), [Pisma Zh.\ Eksp.\ Teor.\ Fiz.\  {\bf 12}, 538 (1970)].
}

\lref\FitzpatrickDM{
  A.~L.~Fitzpatrick and J.~Kaplan,
  ``Unitarity and the Holographic S-Matrix,''
JHEP {\bf 1210}, 032 (2012).
[arXiv:1112.4845 [hep-th]].
}

\lref\CostaDW{
  M.~S.~Costa, J.~Penedones, D.~Poland and S.~Rychkov,
  ``Spinning Conformal Blocks,''
JHEP {\bf 1111}, 154 (2011).
[arXiv:1109.6321 [hep-th]].
}

\lref\OsbornVT{
  H.~Osborn,
  ``Conformal Blocks for Arbitrary Spins in Two Dimensions,''
Phys.\ Lett.\ B {\bf 718}, 169 (2012).
[arXiv:1205.1941 [hep-th]].
}

\lref\LangGE{
  K.~Lang and W.~Ruhl,
  ``Critical O(N) vector nonlinear sigma models: A Resume of their field structure,''
[hep-th/9311046].
}

\lref\KazakovKM{
  D.~I.~Kazakov,
  ``The Method Of Uniqueness, A New Powerful Technique For Multiloop Calculations,''
Phys.\ Lett.\ B {\bf 133}, 406 (1983)..
}

\lref\HogervorstKVA{
  M.~Hogervorst, H.~Osborn and S.~Rychkov,
  ``Diagonal Limit for Conformal Blocks in d Dimensions,''
[arXiv:1305.1321 [hep-th]].
}
\lref\FitzpatrickSYA{
  A.~L.~Fitzpatrick, J.~Kaplan and D.~Poland,
  ``Conformal Blocks in the Large D Limit,''
[arXiv:1305.0004 [hep-th]].
}
\lref\BeemQXA{
  C.~Beem, L.~Rastelli and B.~C.~van Rees,
  ``The N=4 Superconformal Bootstrap,''
[arXiv:1304.1803 [hep-th]].
}
\lref\HogervorstSMA{
  M.~Hogervorst and S.~Rychkov,
  ``Radial Coordinates for Conformal Blocks,''
[arXiv:1303.1111 [hep-th]].
}
\lref\KomargodskiEK{
  Z.~Komargodski and A.~Zhiboedov,
  ``Convexity and Liberation at Large Spin,''
[arXiv:1212.4103 [hep-th]].
}
\lref\FitzpatrickYX{
  A.~L.~Fitzpatrick, J.~Kaplan, D.~Poland and D.~Simmons-Duffin,
  ``The Analytic Bootstrap and AdS Superhorizon Locality,''
[arXiv:1212.3616 [hep-th]].
}
\lref\ElShowkHU{
  S.~El-Showk and M.~F.~Paulos,
  ``Bootstrapping Conformal Field Theories with the Extremal Functional Method,''
[arXiv:1211.2810 [hep-th]].
}
\lref\LiendoHY{
  P.~Liendo, L.~Rastelli and B.~C.~van Rees,
  ``The Bootstrap Program for Boundary CFT$_d$,''
[arXiv:1210.4258 [hep-th]].
}
\lref\PappadopuloJK{
  D.~Pappadopulo, S.~Rychkov, J.~Espin and R.~Rattazzi,
  ``OPE Convergence in Conformal Field Theory,''
Phys.\ Rev.\ D {\bf 86}, 105043 (2012).
[arXiv:1208.6449 [hep-th]].
}
\lref\FriedanJK{
  D.~Friedan, A.~Konechny and C.~Schmidt-Colinet,
  ``Lower bound on the entropy of boundaries and junctions in 1+1d quantum critical systems,''
Phys.\ Rev.\ Lett.\  {\bf 109}, 140401 (2012).
[arXiv:1206.5395 [hep-th]].
}
\lref\ElShowkHT{
  S.~El-Showk, M.~F.~Paulos, D.~Poland, S.~Rychkov, D.~Simmons-Duffin and A.~Vichi,
  ``Solving the 3D Ising Model with the Conformal Bootstrap,''
Phys.\ Rev.\ D {\bf 86}, 025022 (2012).
[arXiv:1203.6064 [hep-th]].
}
\lref\FitzpatrickDM{
  A.~L.~Fitzpatrick and J.~Kaplan,
  ``Unitarity and the Holographic S-Matrix,''
JHEP {\bf 1210}, 032 (2012).
[arXiv:1112.4845 [hep-th]].
}
\lref\FitzpatrickHU{
  A.~L.~Fitzpatrick and J.~Kaplan,
  ``Analyticity and the Holographic S-Matrix,''
JHEP {\bf 1210}, 127 (2012).
[arXiv:1111.6972 [hep-th]].
}
\lref\RychkovET{
  S.~Rychkov,
  ``Conformal Bootstrap in Three Dimensions?,''
[arXiv:1111.2115 [hep-th]].
}
\lref\CostaDW{
  M.~S.~Costa, J.~Penedones, D.~Poland and S.~Rychkov,
  ``Spinning Conformal Blocks,''
JHEP {\bf 1111}, 154 (2011).
[arXiv:1109.6321 [hep-th]].
}
\lref\PolandEY{
  D.~Poland, D.~Simmons-Duffin and A.~Vichi,
  ``Carving Out the Space of 4D CFTs,''
JHEP {\bf 1205}, 110 (2012).
[arXiv:1109.5176 [hep-th]].
}
\lref\DolanDV{
  F.~A.~Dolan and H.~Osborn,
  ``Conformal Partial Waves: Further Mathematical Results,''
[arXiv:1108.6194 [hep-th]].
}
\lref\FitzpatrickIA{
  A.~L.~Fitzpatrick, J.~Kaplan, J.~Penedones, S.~Raju and B.~C.~van Rees,
  ``A Natural Language for AdS/CFT Correlators,''
JHEP {\bf 1111}, 095 (2011).
[arXiv:1107.1499 [hep-th]].
}
\lref\VichiUX{
  A.~Vichi,
  ``Improved bounds for CFT's with global symmetries,''
JHEP {\bf 1201}, 162 (2012).
[arXiv:1106.4037 [hep-th]].
}
\lref\FitzpatrickHH{
  A.~L.~Fitzpatrick and D.~Shih,
  ``Anomalous Dimensions of Non-Chiral Operators from AdS/CFT,''
JHEP {\bf 1110}, 113 (2011).
[arXiv:1104.5013 [hep-th]].
}
\lref\RattazziYC{
  R.~Rattazzi, S.~Rychkov and A.~Vichi,
  ``Bounds in 4D Conformal Field Theories with Global Symmetry,''
J.\ Phys.\ A {\bf 44}, 035402 (2011).
[arXiv:1009.5985 [hep-th]].
}
\lref\PolandWG{
  D.~Poland and D.~Simmons-Duffin,
  ``Bounds on 4D Conformal and Superconformal Field Theories,''
JHEP {\bf 1105}, 017 (2011).
[arXiv:1009.2087 [hep-th]].
}
\lref\RattazziGJ{
  R.~Rattazzi, S.~Rychkov and A.~Vichi,
  ``Central Charge Bounds in 4D Conformal Field Theory,''
Phys.\ Rev.\ D {\bf 83}, 046011 (2011).
[arXiv:1009.2725 [hep-th]].
}
\lref\HellermanQD{
  S.~Hellerman and C.~Schmidt-Colinet,
  ``Bounds for State Degeneracies in 2D Conformal Field Theory,''
JHEP {\bf 1108}, 127 (2011).
[arXiv:1007.0756 [hep-th]].
}
\lref\HeemskerkPN{
  I.~Heemskerk, J.~Penedones, J.~Polchinski and J.~Sully,
  ``Holography from Conformal Field Theory,''
JHEP {\bf 0910}, 079 (2009).
[arXiv:0907.0151 [hep-th]].
}

\lref\RychkovIJ{
  V.~S.~Rychkov and A.~Vichi,
  ``Universal Constraints on Conformal Operator Dimensions,''
Phys.\ Rev.\ D {\bf 80}, 045006 (2009).
[arXiv:0905.2211 [hep-th]].
}
\lref\HellermanBU{
  S.~Hellerman,
  ``A Universal Inequality for CFT and Quantum Gravity,''
JHEP {\bf 1108}, 130 (2011).
[arXiv:0902.2790 [hep-th]].
}
\lref\RattazziPE{
  R.~Rattazzi, V.~S.~Rychkov, E.~Tonni and A.~Vichi,
  ``Bounding scalar operator dimensions in 4D CFT,''
JHEP {\bf 0812}, 031 (2008).
[arXiv:0807.0004 [hep-th]].
}

\lref\Wess{
J.~Wess, ``The Conformal Invariance in Quantum Field Theory'', 
Nuovo Cimento {\bf 6}, 1086 (1960).} 

\lref\PolchinskiDY{
  J.~Polchinski,
  ``Scale And Conformal Invariance In Quantum Field Theory,''
Nucl.\ Phys.\ B {\bf 303}, 226 (1988).
}

\lref\ElShowkGZ{
  S.~El-Showk, Y.~Nakayama and S.~Rychkov,
  ``What Maxwell Theory in D<>4 teaches us about scale and conformal invariance,''
Nucl.\ Phys.\ B {\bf 848}, 578 (2011).
[arXiv:1101.5385 [hep-th]].
}

\lref\FortinHN{
  J.~-F.~Fortin, B.~Grinstein and A.~Stergiou,
  ``Limit Cycles and Conformal Invariance,''
JHEP {\bf 1301}, 184 (2013), [JHEP {\bf 1301}, 184 (2013)].
[arXiv:1208.3674 [hep-th]].
}

\lref\jh{
 C.~Agon,  M.~Headrick, D.~Jafferis, and S.~Kasko,
  to appear.
}

\lref\RivaGD{
  V.~Riva and J.~L.~Cardy,
  ``Scale and conformal invariance in field theory: A Physical counterexample,''
Phys.\ Lett.\ B {\bf 622}, 339 (2005).
[hep-th/0504197].
}

\lref\FortinCQ{
  J.~-F.~Fortin, B.~Grinstein and A.~Stergiou,
  ``Limit Cycles in Four Dimensions,''
JHEP {\bf 1212}, 112 (2012).
[arXiv:1206.2921 [hep-th]].
}

\lref\ZhengBP{
  S.~Zheng,
  ``Is There Scale Invariance in N=1 Supersymmetric Field Theories ?,''
[arXiv:1103.3948 [hep-th]].
}

\lref\ft{
  E.~S.~Fradkin and A.~A.~Tseytlin,
  ``Quantum Equivalence of Dual Field Theories,''
Annals Phys.\  {\bf 162}, 31 (1985).
}

\lref\JackiwVZ{
  R.~Jackiw and S.~-Y.~Pi,
  ``Tutorial on Scale and Conformal Symmetries in Diverse Dimensions,''
J.\ Phys.\ A {\bf 44}, 223001 (2011).
[arXiv:1101.4886 [math-ph]].
}

\lref\SchwarzCN{
  A.~S.~Schwarz,
  ``The Partition Function of Degenerate Quadratic Functional and Ray-Singer Invariants,''
Lett.\ Math.\ Phys.\  {\bf 2}, 247 (1978)..
}

\lref\WittenHF{
  E.~Witten,
  ``Quantum Field Theory and the Jones Polynomial,''
Commun.\ Math.\ Phys.\  {\bf 121}, 351 (1989)..
}

\lref\NakayamaTK{
  Y.~Nakayama,
  ``Comments on scale invariant but non-conformal supersymmetric field theories,''
Int.\ J.\ Mod.\ Phys.\ A {\bf 27}, 1250122 (2012).
[arXiv:1109.5883 [hep-th]].
}

\lref\NakayamaND{
  Y.~Nakayama,
  ``Supercurrent, Supervirial and Superimprovement,''
[arXiv:1208.4726 [hep-th]].
}

\lref\NakayamaFE{
  Y.~Nakayama,
  ``No Forbidden Landscape in String/M-theory,''
JHEP {\bf 1001}, 030 (2010).
[arXiv:0909.4297 [hep-th]].
}

\lref\DeserYX{
  S.~Deser and A.~Schwimmer,
  ``Geometric classification of conformal anomalies in arbitrary dimensions,''
  Phys.\ Lett.\  B {\bf 309}, 279 (1993)
  [arXiv:hep-th/9302047].
}

\lref\IorioAD{
  A.~Iorio, L.~O'Raifeartaigh, I.~Sachs and C.~Wiesendanger,
  ``Weyl gauging and conformal invariance,''
Nucl.\ Phys.\ B {\bf 495}, 433 (1997).
[hep-th/9607110].
}

\lref\KlebanovGS{
  I.~R.~Klebanov, S.~S.~Pufu and B.~R.~Safdi,
  ``F-Theorem without Supersymmetry,''
JHEP {\bf 1110}, 038 (2011).
[arXiv:1105.4598 [hep-th]].
}

\lref\DorigoniRA{
  D.~Dorigoni and V.~S.~Rychkov,
  ``Scale Invariance + Unitarity => Conformal Invariance?,''
[arXiv:0910.1087 [hep-th]].
}

\lref\NakayamaIS{
  Y.~Nakayama,
  ``A lecture note on scale invariance vs conformal invariance,''
[arXiv:1302.0884 [hep-th]].
}

\lref\Theisen{
  S.~Theisen, unpublished
}

\lref\SchwimmerZA{
  A.~Schwimmer and S.~Theisen,
  ``Spontaneous Breaking of Conformal Invariance and Trace Anomaly Matching,''
Nucl.\ Phys.\ B {\bf 847}, 590 (2011).
[arXiv:1011.0696 [hep-th]].
}

\lref\ElvangST{
  H.~Elvang, D.~Z.~Freedman, L.~-Y.~Hung, M.~Kiermaier, R.~C.~Myers and S.~Theisen,
  ``On renormalization group flows and the a-theorem in 6d,''
JHEP {\bf 1210}, 011 (2012).
[arXiv:1205.3994 [hep-th]].
}

\lref\DeserCA{
  S.~Deser and A.~Schwimmer,
  ``Gauge field improvement, form - scalar duality, conformal invariance and quasilocality,''
Int.\ J.\ Mod.\ Phys.\ B {\bf 8}, 3741 (1994).
[hep-th/9404183].
}

\lref\DolanUT{
  F.~A.~Dolan and H.~Osborn,
  ``Conformal four point functions and the operator product expansion,''
Nucl.\ Phys.\ B {\bf 599}, 459 (2001).
[hep-th/0011040].
}

\lref\boaz{
 F.~Baume, B.~Keren-Zur, R.~Rattazzi, and L.~Vitale,
  to appear.
}

\lref\AldayMF{
  L.~F.~Alday and J.~M.~Maldacena,
  ``Comments on operators with large spin,''
JHEP {\bf 0711}, 019 (2007).
[arXiv:0708.0672 [hep-th]].
}

\lref\Migdal{
  A. Migdal,
  ``Ancient History of CFT,''
The 12th Claude Itzykson Meeting (2007).
}


\rightline{WIS/10/13 SEP-DPPA}
\vskip-20pt
\Title{
} {\vbox{\centerline{On Scale and Conformal Invariance in Four Dimensions}}}
\medskip

\centerline{\it Anatoly Dymarsky,$^{a}$ Zohar Komargodski,$^{b}$ 
Adam Schwimmer,$^{b}$ and Stefan Theisen\ $^{c}$}

\bigskip

\centerline{$^a$  DAMTP, University of Cambridge, Cambridge, CB3 0WA, UK}
\centerline{$^b$ Weizmann Institute of Science, Rehovot
76100, Israel}
 \centerline{$^c$
Max-Planck-Institut f\"ur Gravitationsphysik, Albert-Einstein-Institut, 14476 Golm, Germany}

\smallskip

\vglue .3cm
\bigskip
\bigskip
\bigskip
\noindent

We study the implications of scale invariance in four-dimensional quantum field theories.
Imposing unitarity, we find that infinitely many matrix elements vanish in a suitable kinematical configuration. This vanishing is a nontrivial necessary condition for conformality. We provide an argument why this is expected to be a sufficient condition as well, thereby linking scale and conformal invariance in unitary theories. We also discuss possible exceptions to our argument.

\Date{September 2013}

\newsec{Introduction and Summary}

Consider Poincar\'e-invariant Quantum Field Theories in $d$ space-time dimensions. A special class of such  theories contains the theories  that have no intrinsic mass scale. In other words, (at least intuitively) the correlation functions can be chosen to behave homogeneously if we rescale all the distances. The symmetry group in this case contains 
 \eqn\scalesymmetrygroup{ISO(d-1,1)\rtimes \IR^+~,}
where $ISO(d-1,1)$ is  the Poincar\'e group (in Minkowski space) and $\IR^+$ is generated by dilatations, $\hat D$. Lorentz transformations are invariant under dilatations, while the momentum generators carry charge 1. Surprisingly, in most of the studied examples, the symmetry group~\scalesymmetrygroup\ is in fact enhanced to the conformal group
\eqn\scalesymmetrygroupi{SO(d,2)~.}
The additional generators in \scalesymmetrygroupi\ compared with \scalesymmetrygroup\ lead to numerous constraints on the critical exponents  and impose important restrictions on general $n$-point correlation functions.\foot{Recently, there has been a spur of activity analyzing the constraints of $SO(d,2)$ via the bootstrap equations~\refs{\FGG, \PolyakovGS, \BPZ} with new analytic and numerical tools, see e.g.~\refs{\DolanUT\AldayMF\RattazziPE\HellermanBU\HeemskerkPN\ElShowkHT\LiendoHY\FitzpatrickYX\KomargodskiEK-\BeemQXA}.} For instance, conformal symmetry fixes the correlation function of any three local operators up to finitely many real numbers, while scale invariance alone allows for undetermined functions. 

We will refer to the models that are invariant under~\scalesymmetrygroup\ as scale-invariant field theories (SFTs) while the models invariant under~\scalesymmetrygroupi\ will be called conformal field theories (CFTs). Of course, conformal theories are scale invariant. The converse statement is the subject of this paper.\foot{The reader interested in the history of this problem can consult, for instance,~\Migdal.}

Let us define the problem more precisely. We start from the class of theories which have a conserved, symmetric, local energy momentum tensor 
\eqn\emtensor{T_{\mu\nu}=T_{\nu\mu}~,\qquad \del^\mu T_{\mu\nu}=0~.}
If there exists a local operator $V_\nu$ (referred to as the ``virial current") such that 
\eqn\virial{T_\mu^\mu=\del^\nu V_\nu~,}
one can construct a conserved scale 
(or dilatation) current 
\eqn\dilationcurrent{S_\mu=x^\nu T_{\mu\nu}-V_\mu~,\qquad \del^\mu S_\mu=0~,}
and a dilatation charge $\hat{D}=\int d^{d-1}x\, S_0$.
The converse is also true: if there exists a local conserved scale current then it takes the form~\dilationcurrent\ (see~\refs{\Wess,\CallanZE,\ColemanJE} for general background on the subject). 
If, furthermore, there exists a local operator $L_{\mu\nu}$ such that
$V_\mu=\partial^\nu L_{\nu\mu}$ and correspondingly
\eqn\conformalcondition{
T^\mu_\mu=\p^\rho\p^\sigma L_{\rho\sigma}\,,}
one can construct the conserved conformal currents   
\eqn\conformalcurrent{
K^{\mu\nu}=(2 x^\nu\,x_\rho-x^2\delta^\nu_\rho)T^{\rho\mu}
-2\, x^\nu\,V^\mu+2\,L^{\mu\nu}~,\qquad \del_\mu K^{\mu\nu}=0~,}
and the theory is invariant under the full conformal group \scalesymmetrygroupi. Equivalently, one notices that if~\conformalcondition\ is satisfied then there exists an ``improved'' symmetric, conserved, and traceless $T_{\mu\nu}$ which is constructed of the original stress-energy tensor appropriately shifted by two-derivative terms acting on $L_{\mu\nu}$.\foot{Another equivalent description of the conditions for scale vs. conformal invariance can be given by coupling the theory to curved space  and demanding Weyl invariance, see, for example,~\IorioAD.}

In four dimensions, if a SFT is unitary, the condition that the theory is conformal~\conformalcondition\ can be simplified (this follows from the unitarity bound on operator dimensions~\GrinsteinQK, see appendix A) to $V_\mu=\del_\mu L$, i.e.,
\eqn\conformalconditioni{T_\mu^\mu=\square L~.} 
Equation~\conformalconditioni\ is a necessary and sufficient condition for a unitary scale invariant theory to be conformal. The goal  of this paper is to argue that unitarity and scale invariance imply \conformalconditioni\ in four dimensions (under some additional assumptions and caveats that will be specified in the following).

In $d=2$ the situation is conceptually simpler. A hypothetical $L$ must have dimension~$0$ and under some technical assumptions it can be ruled out, therefore, leaving $T_\mu^\mu=0$ as the only possible condition satisfied by a stress-energy tensor in a conformal theory. In other words, roughly speaking, the origin of the relative simplicity of the problem in $d=2$ is that no improvements of the energy-momentum tensor are possible.  It was shown in~\PolchinskiDY\ that the traceless-ness of the stress-energy tensor follows in $d=2$ from scale invariance and unitarity (again, under some technical assumptions). The argument
revolves around the two-point function of the energy momentum tensor,
and  is reminiscent of the proof of the $c$-theorem~\ZamolodchikovGT. By manipulating the two-point function one can show that, assuming scale invariance,  one obtains $\langle T_\mu^\mu(x) T_\nu^\nu(0)\rangle=0$ thus proving $T_\mu^\mu=0$.

One cannot hope to repeat an argument of this type in higher dimensions simply because there are  nontrivial improvements of the stress-energy tensor in $d>2$ CFTs. More concretely, it is not true that scale invariance implies (even when combined with unitarity) $T_\mu^\mu=0$.
Instead, one must show that the theory has a local operator $L_{\mu\nu}$ such that~\conformalcondition\ is satisfied (for unitary theories we need to show \conformalconditioni\ instead). It could be that $L_{\mu\nu}=0$, but it is not necessarily the case.

In perturbation theory around some Gaussian point, the set of local operators is well understood and one can check explicitly whether there is or there is not  an obstruction for solving~\conformalcondition\ in any given model. However, to venture into 
the non-perturbative regime one would have to construct the operator $L_{\mu\nu}$ (or $L$) formally. 

Recently, there has been progress on the problem of scale versus conformal invariance in $d>2$.  Using tools similar to the derivation of the $a$-theorem in $d=4$~\refs{\KomargodskiVJ,\KomargodskiXV}, Luty, Polchinski, and Rattazzi~\LutyWW\ have provided an argument that unitary four-dimensional  scale-invariant models defined in the vicinity of a CFT must be conformal. This is consistent with a very detailed study of a large class of perturbative models~\refs{\PolchinskiDY,\DorigoniRA,\FortinCQ,\FortinHN}. We would also like to note that recently several authors have proposed arguments that are particular to supersymmetric theories~\refs{\AntoniadisGN\ZhengBP\NakayamaTK-\NakayamaND} and to theories with weakly-coupled holographic duals~\NakayamaFE. Further references on the subject can be found  in~\refs{\DorigoniRA,\JackiwVZ,\NakayamaIS}.

Let us describe the main idea of this paper and defer for a while the discussion of some crucial technical details.  Suppose we have an operator $\CO$ and we want to prove that there exists a local operator $L$ such that $\CO=\square L$. Of course we can always solve for $L$ formally in terms of a nonlocal integral over space-time, but we want to understand under which conditions $L$ would be local. A natural way to tackle this problem is  as follows. We can deform the action by coupling $\CO$ to a classical source $\delta S= \int d^dx\, J(x)\, \CO(x)+O(J^2)$. We can then consider the generating functional of connected diagrams $W[J(x)]$, and its Fourier-transformed version $W[J(p)]$. 
Now let us study  $W[J(p)]$   with null momenta (i.e.~zero momenta squared) for the source (one may need to regulate the theory in the infrared to make sure this object exists): 
\eqn\source{{\delta^m W[J(p)]\over\delta J(p_1)\delta J(p_2)\dots\delta J(p_m)}\biggr|_{J=0;\ p_1^2=p_2^2=\dots=p_m^2=0}\equiv \CA_m(p_1,\dots,p_m)~.}
Suppose that the theory is unitary and that we managed to show that for all $m>2$, $\CA_m(p_1,\dots,p_m)=0$. Then from unitarity it follows that all the  ``on-shell'' matrix elements connecting insertions of $J$ to the physical theory vanish. So, on the one hand, all the matrix elements interpolating between $J$ (with null momenta) and our theory vanish, but, on the other hand, the coupling 
$\int d^dx\, J(x)\, \CO(x)+O(J^2)$ seems nontrivial. This can be consistent if the coupling $\int d^dx\, J(x)\, \CO(x)+O(J^2)$ is trivial on-shell, i.e. $\CO=\square L$ for some local $L$ (in particular, $L$ may vanish).\foot{Another option is  that there are no connected diagrams whatsoever (aside from the two-point function), i.e., $\CO$ is a generalized free field. We will comment on this unlikely possibility in the main text, although we will not be able to rule it out. For simplicity, we ignore this possibility in this section.} 

Clearly, $\CO=\square L$ is a sufficient condition for the vanishing of the above-mentioned matrix elements, but is it necessary? The following argument suggests that the answer is affirmative, at least in a class of sufficiently well-behaved QFTs.

Let us  think of $J(x)$ as a dynamical field associated with a massless elementary particle very weakly coupled to the original theory via $\int d^d x\, J(x)\,\CO(x)$ (one can include an arbitrarily small coupling constant in front of this term in the action). Then, the amplitudes~\source\ can be interpreted as the S-matrix elements describing scattering of on-shell $J$ quanta. This interpretation is correct to leading order in the coupling between $J(x)$ and $\CO(x)$ since internal lines of the $J$ particle need not be included at the leading order. 

Now, a familiar observation in the theory of the S-matrix is that it  is invariant under changes of variables, and moreover, the S-matrix is trivial only if the theory is free after some change of variables. Here we have a variant of this situation: all the matrix elements interpolating between insertions of $J$ (with null momenta) and our theory vanish. It is then expected that to leading order we can remove the interaction $\int d^dx\, J(x)\, \CO(x)+O(J^2)$. This can be done only if $\CO=\square L$ for some local $L$. In this case we can redefine our dynamical field $J$ by shifting it by $L$, thereby removing the small coupling between our fiducial propagating particle and the theory.\foot{In detail, if we have the action $\int d^dx\left(J\square J+J(x)\CO(x)+\cdots\right)$~, and if $\CO=\square L$ then we can redefine $J'=J+\half L+\cdots$ and remove the coupling of $J$ to the composite operator in the theory. We will see in examples that, in the process of doing such field redefinitions, the seagull terms $\CO(J^2)$ disappear as well (as they should) to the required order in the coupling between $J$ and the original theory. } 

While this property of the S-matrix is very intuitive, we are not aware of a general proof. Since this argument plays an important role in our construction, our paper should only be viewed as a physical explanation of why  unitary scale invariant field theories are conformal, but perhaps not as a mathematical theorem. 


In fact, in section~4 we will see that the example of the free two-form theory with noncompact
gauge symmetry in four dimensions is inconsistent with this intuitive-sounding
assertion about the S-matrix. (However, the free two-form theory with compact gauge
symmetry is perfectly consistent with our assertions.) The key is that, in flat space, the
free two-form theory is utterly indistinguishable from the ordinary non-compact scalar,
which is of course conformal. The Hilbert spaces of these theories agree.
 
The free two-form theory with non-compact gauge symmetry can be thought of as the ordinary non-compact free scalar theory from which we remove some of the local operators, while not modifying other correlation functions and not modifying the Hilbert space. This is the source of the obstruction preventing the free two-form theory from being conformal. For nontrivial interacting QFTs, we do not expect that one can have two completely equivalent and consistent theories in flat space, differing in their content of {\it local} operators. In general, if one tries to remove some set of operators from the theory without modifying anything else, the theory becomes inconsistent (this can be seen explicitly in many  two-dimensional examples).

To investigate scale invariance in this vein it is natural to couple a source, $\Psi(x)$, to the trace of the energy momentum tensor $\delta S\sim  \int d^dx\  \Psi(x) T_\mu^\mu(x)+\cdots$, where the~$\cdots$ stand for various seagull terms -- higher order terms in $\Psi$ which ensure diffeomorphism invariance beyond linear level. (We think of $1+\Psi(x)$ as the scale factor of a conformally flat background metric field. There are several different definitions in the literature of the trace of the energy momentum tensor. The difference is just in the structure of contact terms. The convention we follow is explained in section~3.)
We will show that in $d=4$, assuming unitarity, all the amplitudes~\source\ for $m>2$ vanish in the scale invariant theory. We  utilize the power of anomalies and the particular structure of counter-terms in four dimensions to establish this vanishing theorem. 
This means (with the qualification mentioned above)  that the trace of the stress-energy tensor is $T_\mu^\mu=\square L$ for some local $L$ (which could be zero). Hence, our unitary SFT is in fact a CFT.  

The analysis of the case of $m=4$ in~\source, i.e.~$2\rightarrow 2$ scattering of the external sources, was conducted in~\LutyWW, where the triviality of this amplitude  was shown. It was then argued that $\CA_4=0$ implies that unitary SFTs which are perturbatively close to CFTs must, in fact, be conformal. (Loosely speaking, when one stays perturbatively close to a CFT, the flow can be described using the leading-order $\beta$-functions that appear in conformal perturbation theory. The vanishing of $\CA_4$ implies that these $\beta$-functions vanish, hence, the nearby theory is conformal.) Our extension of this analysis to arbitrary scattering processes involving  the external source $\Psi(x)$ allows us to explore the problem beyond the perturbative regime.

This paper is organized as follows. 
In section 2 we discuss in detail the definition of scale invariance and  also review various issues related to improvement terms and the current algebra in such theories. In section 3 we present the main argument  relating scale invariance and conformal invariance. In section 4 we discuss several simple examples, and one subtle example. 
A technical discussion of the case of non-diagonal action of the dilatation operator $\hat{D}$ is given in appendix~A, where we also justify~\conformalconditioni\ in the most general setting. In appendix~B we present new sum rules for RG flows between CFTs. We apply those to flows  of the type CFT-SFT-CFT, which can spend arbitrarily long RG time near the SFT. 

\newsec{A Closer Look at Scale  Invariance}

In this section we discuss in detail the definition of scale invariance. Unless specified otherwise, our discussion is general, i.e.,~it does not assume $d=4$ or unitarity.

\subsec{Aspects of Current Algebra}
To begin, we would like to investigate how the scaling charge $\hat D=\int d^{d-1}x\, S_0$ acts on local operators such as the stress-energy tensor and the virial current. (In fact, for many of the results in this subsection we do not need to assume the existence of the scale current, but just of the scale charge.) The most general algebra consistent with conservation and symmetry of the stress-energy tensor is (see for instance~\PolchinskiDY)
\eqn\currentzero{i[\hat D, T_{\mu\nu}]=x^\rho\del_\rho T_{\mu
\nu}+d\, T_{\mu\nu}+\del^\rho\del^\sigma Y_{\rho\mu ; \sigma\nu}~.}
The operator $Y_{\rho\mu ; \sigma\nu}$ has the same symmetries as the Riemann tensor, i.e.~it is anti-symmetric in $\rho\mu$, anti-symmetric in  $\sigma\nu$, and symmetric under the exchange of the two pairs. The coefficient in front of the second term is fixed to be $d$ by requiring $i[\hat D,H]=H$. The formula~\currentzero\ holds even if $\hat D$ is spontaneously broken in the vacuum state. Suppose for a moment that $\hat D$ is a symmetry of the vacuum. Then, in the absence of the operator $Y$ on the right hand side of~\currentzero, the Ward identities are the naive ones expected in scale invariant QFTs, so that correlation functions obey homogenous scaling relations. We call the current algebra without the operator $Y$ ``canonical.'' In the presence of the operator $Y$, dilatations are realized nontrivially, mixing  correlation functions of the energy-momentum tensor with other correlation functions containing the operator $Y$. 

From~\virial\  and $[\hat D, \hat D]=0$ we deduce the current algebra of the virial current
\eqn\currenti{i[\hat D, V_{\mu}]=x^\rho\del_\rho V_{\mu}
+(d-1) V_{\mu}+\del^\nu Y_{\mu\nu}+\CC_\mu~.} 
In the above, $Y_{\mu\nu}=\eta^{\rho\sigma}Y_{\rho\mu ; \sigma\nu}$ and $\CC_\mu$ is a conserved current that further satisfies $\int d^{d-1}x \ \CC_0=0$.

We denote all the operators in the theory of the type appearing  in~\currentzero\ by $Y^I_{\rho\mu ; \sigma\nu}$. They necessarily satisfy a current algebra of the  following form: 
\eqn\currentii{i[\hat D, Y^I_{\rho\mu ; \sigma\nu}]=x^\lambda\del_\lambda Y^I_{\rho\mu ; \sigma\nu}+\Gamma^{I}_{J} Y^J_{\rho\mu ; \sigma\nu}~.}
By convention, the specific operator appearing in~\currentzero\ is denoted $Y_{\rho\mu ; \sigma\nu}\equiv y_I Y^I_{\rho\mu ; \sigma\nu}$.

The matrix $\Gamma^{I}_{J}$ can always be brought to its canonical Jordan form. The diagonal elements of the Jordan form of $\Gamma^{I}_{J}$ are the generalized dimensions. If we further assume that all the correlation functions in the theory decay at long distances,  then all the generalized dimensions must be positive, except for the unit operator which has dimension zero. 

Since the energy-momentum tensor is not unique, we can try to remove the offensive term on the right hand side of~\currentzero\ by an improvement transformation.  The most general possible improvement transformation is 
\eqn\improvement{T_{\mu\nu}\rightarrow T_{\mu\nu}+w_I\partial^\rho\partial^\sigma Y^I_{\rho\mu ; \sigma\nu}~,}
where the $w_I$ are arbitrary coefficients. The new $T_{\mu\nu}$ will satisfy the current algebra~\currentzero\ with $y'_I=y_I+(\Gamma_I^J+(2-d)\delta_I^J)w_J$. We can therefore transform~\currentzero\ to the canonical current algebra (i.e.~$y'_I=0$) unless there are operators $Y^I$ with generalized dimension $d-2$. (The positivity of generalized dimensions rules out such a possibility in $d=2$, but in $d>2$ the action of dilatation may not be canonical if such operators with generalized dimension $d-2$ exist.\foot{We thank Y.~Nakayama for a discussion on this issue.})

Consider the space of local operators at the origin. $\hat D$ provides a natural linear map on this space $\cdot\rightarrow [\hat D, \cdot]$. We see that if the last term in~\currentzero\ is not removable, then $\hat D$ is non-diagonalizable. 

Let us briefly comment on the current algebra in unitary theories. In four dimensions, if our SFT is unitary one can show (see Appendix~A) that the only allowed $Y^I_{\rho\mu;\sigma\nu}$ of generalized dimension $2$ must  be a scalar $Y_{\rho\mu ; \sigma\nu}=(\eta_{\rho\sigma}\eta_{\mu\nu}-\eta_{\rho\nu}\eta_{\mu\sigma})Y$ or a conserved tensor.
Thus, the most general current algebra for the trace of the energy-momentum tensor in unitary scale-invariant four-dimensional theories takes the form 
\eqn\currentiv{i[\hat D, T_{\mu}^\mu]=x^\rho\del_\rho T_{\mu}
^\mu+d\,  T_{\mu}^\mu+\square Y~.}

We will henceforth make the assumption (this assumption is implicit in most of the literature on the subject) that $\hat D$ is diagonalizable, and in particular, the current algebra can be chosen to be canonical.  As we explain in appendix~A, this assumption is not necessary to derive our main results. However, making this assumption somewhat simplifies the presentation.

\subsec{The Background Functional and SFT Anomalies}

An alternative language to the current algebra is the background functional method. This has the advantage of allowing to classify the anomalies easily.  Let us  imagine coupling our scale invariant theory to a background metric  $g_{\mu\nu}$ and a vector field $C_\mu$, such that under an infinitesimal deformation of $g_{\mu\nu}$ and $C_\mu$
\eqn\actionchange{\delta S=\int d^d x\, \sqrt g \left(\half T^{\mu\nu}\delta g_{\mu\nu}+V^\mu\delta C_\mu\right)~.}
As long as our current algebra is canonical (\currentzero,\currenti~without the $Y$ and $\CC_\mu$ terms), the generating functional $W[g_{\mu\nu},C_\mu]$ is invariant (up to anomalies) under \eqn\localweyl{\delta g_{\mu\nu}=2\sigma g_{\mu\nu}~,\qquad \delta C_\mu=\del_\mu \sigma~.}
(If the current algebra is non-canonical, to realize Weyl invariance, the generating functional $W$ must also depend on additional background fields that couple to the operators $Y$. The modified transformation rules in the presence of such background fields can be worked out as in~\refs{
\OsbornGM,\boaz}.) 

We can now classify $c$-number anomalies (for simplicity, we ignore parity-violating anomalies).  We are looking for  local functionals $\CA( g_{\mu\nu}, C_\mu)$ obeying the Wess-Zumino consistency condition. From the definition of $\CA(g_{\mu\nu}, C_\mu)$   
\eqn\anomaly{\delta_{\sigma(x)} W[g_{\mu\nu}, C_\mu]=\int d^dx\,\sqrt g\,\sigma(x) \CA(g_{\mu\nu}, C_\mu) ~,}
we get a nontrivial constraint on $\CA(g_{\mu\nu}, C_\mu)$ by imposing  $[\delta_{\sigma_1},\delta_{\sigma_2}]=0$. 
One finds that in four dimensions there are  four anomalies in total \LutyWW:
\eqn\anomalies{\CA( g_{\mu\nu}, C_\mu)=aE_4+cW^2+b\left(R+6\nabla\cdot C-6C^2\right)^2+\gamma \del_{[\mu}C_{\nu]}\del^{[\mu}C^{\nu]} ~.}
The $a$ and $c$ anomalies satisfy the Wess-Zumino consistency condition in the same way as for the usual trace anomaly. The anomalies proportional to $b,\gamma$ are particular to SFTs. The anomalies $b,\gamma$ are both ``type-B'' in the sense that they satisfy the Wess-Zumino consistency condition trivially, because they are Weyl invariant. All the anomalies in~\anomalies\ are genuine, i.e.~they cannot be removed by adding a local (diffeomorphism-invariant) term to $W[g_{\mu\nu},C_\mu]$.

\newsec{Probing SFTs with Renormalization Group Flows}

\subsec{Convergent Dispersion Relations}

Imagine any RG flow of the type depicted in figure 1. In the UV the theory is some SFT (which could be a CFT) and we flow to a gapped phase.\foot{The main argument can be extended to the case when the infrared is a CFT.} The crossover scale is denoted by $M$. We can couple the theory to a background metric $g_{\mu\nu}$ in a coordinate-invariant fashion. Since the theory in the infrared is gapped (it could have some topological degrees of freedom but those are inconsequential for our discussion) the low energy effective action is  a {\it local} functional of the background metric.  This local functional can be expanded in derivatives. Up to four derivatives, discarding total derivative terms, we find 
\eqn\lemetric{S_{IR}[g_{\mu\nu}]=\int d^4x\sqrt g \left(\Lambda+\frak{a}  R + \frak{b}R^2+\frak{c}W^2+\CO(\del^6)\right)~.}
Here $W$ is the Weyl tensor. The two remaining contractions with four derivatives, one parity even and one parity odd, are ``total derivatives'' corresponding to the Euler and Pontryagin topological invariants.

\ifig\figten{We consider a renormalization group flow from some SFT to a gapped theory.} {\epsfxsize1.5in\epsfbox{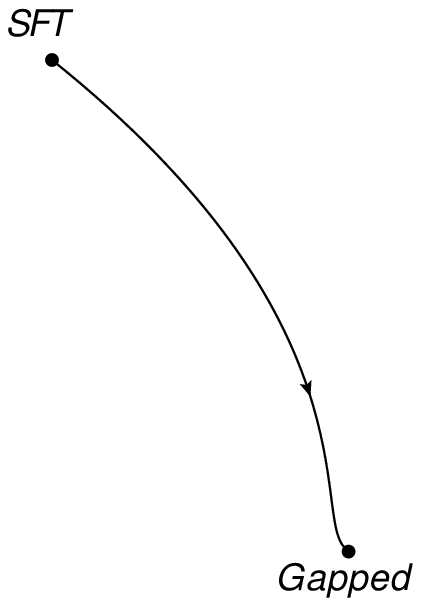}}

All the coefficients in the effective action~\lemetric, $\Lambda, \frak{a},\frak{b},\frak{c}$ are  renormalization scheme dependent. For example, the value of the cosmological constant in the infrared is famously incalculable in quantum field theory. These low energy coefficients do not have a universal meaning  precisely  because the corresponding local terms~\lemetric\ (they all have dimensions $\leq 4$) can be added as counterterms in the UV. 

Now let us consider metrics of the type $g_{\mu\nu}=(1+\Psi)^{2}\eta_{\mu\nu}$ such that $\square \Psi=0$. The UV counter-term corresponding to the coefficient $\frak{c}$ disappears because such metrics are conformally flat. In addition, the structures $\int d^4x\sqrt g R^2$, $\int d^4x\sqrt g R$ do not play a role  because  $R[\Psi^2\eta_{\mu\nu}]\sim \square \Psi=0$. However, the cosmological constant term remains.

Therefore, if we consider the  low energy action as a functional of  $\Psi$ and denote the fourth derivative with respect to $\Psi$ (always evaluating the derivatives at $\Psi=0$) by $\CA_{4}$, then we get an object in which the only allowed scheme ambiguity does not carry momentum dependence (since it comes from the cosmological constant). Therefore, if we take any derivative of $\CA_4$ with respect to momentum we get a perfectly well-defined observable in Quantum Field Theory. Note that derivatives with respect to $\Psi$ are very closely related to the usual definition of correlators of the trace of the energy-momentum tensor. The difference from other conventions is at most by the choice of contact terms. Here the definition of $\langle T_\mu^\mu T_\nu^\nu\cdots\rangle$ that follows from taking derivatives with respect to $\Psi$ is the most convenient one for our purposes.

 We refer to $\CA_4$ as the $2\rightarrow2$ dilaton scattering amplitude for the following reason:
 we can introduce a kinetic term for $\Psi$ with a very large coefficient, and tune the infrared cosmological constant to zero. Then $\CA_4$ becomes a genuine $2\rightarrow2$ scattering amplitude of a massless particle $\Psi$. The fact that the kinetic term has a very large coefficient guarantees that we can ignore diagrams where $\Psi$ appears in internal legs.

Similarly, we can define $\CA_n$ as the $n$-th derivative of the generating functional with respect to $\Psi$. This is again guaranteed to be free of renormalization scheme ambiguities up to a momentum-independent constant. Upon tuning the infrared cosmological constant to zero, $\CA_n$ can be interpreted as a massless $n$-dilaton scattering amplitude.

We can use this understanding of the structure of counter-terms to place bounds on the imaginary part of our scattering processes. Let us begin with $\CA_4$ in the ``on-shell'' kinematics $p_i^2=0$ and further restrict to the forward limit, $p_1=-p_3$, $p_2=-p_4$ and denote $s=(p_1+p_2)^2$ (we are using the mostly minus signature). In the forward kinematics the amplitude depends only on the variable $s$. 

Then we have the usual dispersion relation between the amplitude and its imaginary part 
\eqn\dispersionrelation{\CA_4(s)={1\over \pi} \int\limits_{-\infty}^\infty ds' {\Im \CA_4(s')\over s-s'}+{\rm subtractions}~,}
where $\Im\CA_4(s)=-\Im\CA_4(-s)$ on the real axis.
Since the only allowed {\it divergent} subtraction is momentum independent (cosmological constant), the second derivative of~\dispersionrelation\ should be convergent, hence, we infer that 
\eqn\dispersionbound{\lim_{s\rightarrow\infty}\Im \CA_4(s)<s^2~.}
So far, we have only used the fact that divergent counter-terms for $\CA_4$ are absent and this led to our bound~\dispersionbound. However, there may be finite counter-terms which are momentum dependent. We would not need to discuss those in detail here.\foot{Such finite counter-terms may be associated to anomalies. To have an example in mind of how such finite counter-terms arise, consider dimensional regularization (which although is defined only in perturbation theory, serves us well to demonstrate  how finite counter-terms can arise). Then we have a new counter-term in $4-\epsilon$ dimensions, which is essentially the dimensionally continued Euler density. This is multiplied by $1\over \epsilon$, i.e. the usual logarithmic divergence. The integral of the Euler density vanishes as we take $\epsilon\rightarrow0$ but this is compensated by $1\over \epsilon$ to leave behind a finite counter-term which cannot be wrritten as a local functional in four dimensions. See~\DeserYX\ for further details. A more detailed discussion of these finite counter-terms for flows of the type CFT-SFT-CFT is given in Appendix B, where we utilize the power of infinitely many new sum rules for the difference between the 
$a$-anomalies of the UV and IR CFTs.}

Note that a behavior at large $s$ of the form $\Im\CA_4\sim M^\epsilon s^{2-\epsilon}$ with $\epsilon>0$ is allowed. (The first derivative of $\int_{-\infty}^\infty ds' {\Im \CA_4(s')\over s-s'}$ at $s=0$ automatically vanishes due to the fact that $\Im\CA_4$ is odd. Hence, the first derivative  does not place constraints on the large $s$ behavior.)  One can interpret $\epsilon$ as being related to the dimension of the relevant operator starting the flow of figure~1.

It is straightforward to repeat these ideas for all the  $\CA_{2n}$. Let us arrange  on-shell $p_i^2=0$ forward kinematics  $p_1=-p_{n+1}$, $p_2=-p_{n+2}$,...,$p_n=-p_{2n}$ and define $s_{ij}=(p_i+p_j)^2$. 
At high energies the theory is arbitrarily close to the scale invariant UV fixed point. 
Then, by dimensional analysis\foot{Dimensional analysis holds because the dilation operator assigns the well-defined dimension $d$ to the energy-momentum tensor, as discussed in the previous section.} 
\eqn\ansatzSFT{M^2 \ll s_{ij}, \lambda s_{ij}:\qquad \Im\CA_{2n}(\lambda s_{ij})= \lambda^2 \CF_{2n}(s_{ij})~.}
However, if the function $\CF_{2n}(s_{ij})$ is non-vanishing, such a behavior leads to a contradiction with the absence of counter-terms in the ultraviolet. Hence, the high-energy limit of $\CA_{2n}$ is such that \eqn\vanishingcf{\CF_{2n}=0~.} This result will be crucial in what follows. (One can also derive~\dispersionbound\ and~\vanishingcf\ using the classification of anomalies in subsection 2.2.)

The possible counter-terms~\lemetric\ are limited only by diffeomorphism invariance and power counting. Hence, had we been able to show that the amplitudes $\CA_{2n}$ exist in the SFT itself (i.e. the forward limit is non-singular in the unregulated theory), then we would have established~\vanishingcf\ even in theories that do not admit an appropriate relevant perturbation. In examples that we considered in detail, the forward limit exists even without an infrared regulator. Hence, perhaps the assumption about the existence of the flow in figure~1 can be relaxed. Hereafter, the assumption that the theory can be regulated in the infrared is often implicit.

All the statements in this subsection hold in any renormalization group flow of the type appearing in figure~1, {\it regardless of unitarity}. 

\subsec{Lessons for SFTs}

The results of the previous subsection can be summarized by saying that the imaginary parts of the on-shell, forward scattering amplitudes $\CA_{2n}$  at the SFT itself  vanish. 
Using unitarity, we will now show that the vanishing of the imaginary parts of the scattering amplitudes leads to the fact that $\Psi$ is entirely decoupled from the SFT. This is true in the sense that all the connected matrix elements connecting $m>1$ insertions of (on-shell) $\Psi$ and any state in our theory vanish.

First, we begin with the special case of 2-2 scattering. In the previous subsection we have shown that the imaginary part of $\Im\CA_4$ vanishes identically at the SFT. In unitary theories we can employ the optical theorem which implies that  the two-dilaton matrix elements $\langle \Psi(p_1)\Psi(p_2)| Anything\rangle$ vanish for arbitrary null $p_1$ and $p_2$.\foot{While in non-unitarity theories it is still true that 
$\Im\CA_4$ vanishes identically at the SFT, one cannot conclude from this that the source $\Psi(x)$ is decoupled. Some matrix elements with negative-norm states can cancel against matrix elements with states of positive norm.} This decoupling at the level of $2\rightarrow 2$ scattering is precisely the result of~\LutyWW. 

We can generalize the argument to any $\CA_{2n}$. It is useful to start from $\CA_6$, i.e. $3\rightarrow 3$ scattering, see figure 2. In the previous subsection we have shown that the total imaginary part of this amplitude vanishes in the SFT  (i.e. $\CF_6=0$, see~\vanishingcf). A technical complication is that unlike the case of $2\rightarrow 2$ scattering, there exist non-manifestly positive unitarity cuts (see~\book\ for a pedagogical exposition of these ideas\foot{The fact that there appear non-positive combinations of matrix elements in $k\rightarrow k$ scattering for $k>2$ has also played a crucial role in the analysis of~\ElvangST}). For example, consider the $2\rightarrow 4$ cut in figure 2. The crucial point is that since we have already shown that all the two-particle cuts vanish, such cuts are absent from figure 2. Thus only the first diagram on the right hand side of figure 2 is present. Using the fact that the first term on the right hand side of figure 2 is positive, we conclude that in the SFT the matrix elements 
$\langle \Psi(p_1)\Psi(p_2)\Psi(p_3)| Anything\rangle=0$ for arbitrary null $p_1$, $p_2$, $p_3$.
\ifig\figten{The optical theorem for 3-3 scattering. The first diagram on the right hand side corresponds to a three-particle cut and the second diagram  to a two-particle cut. The latter is not manifestly positive.} {\epsfxsize5.0in\epsfbox{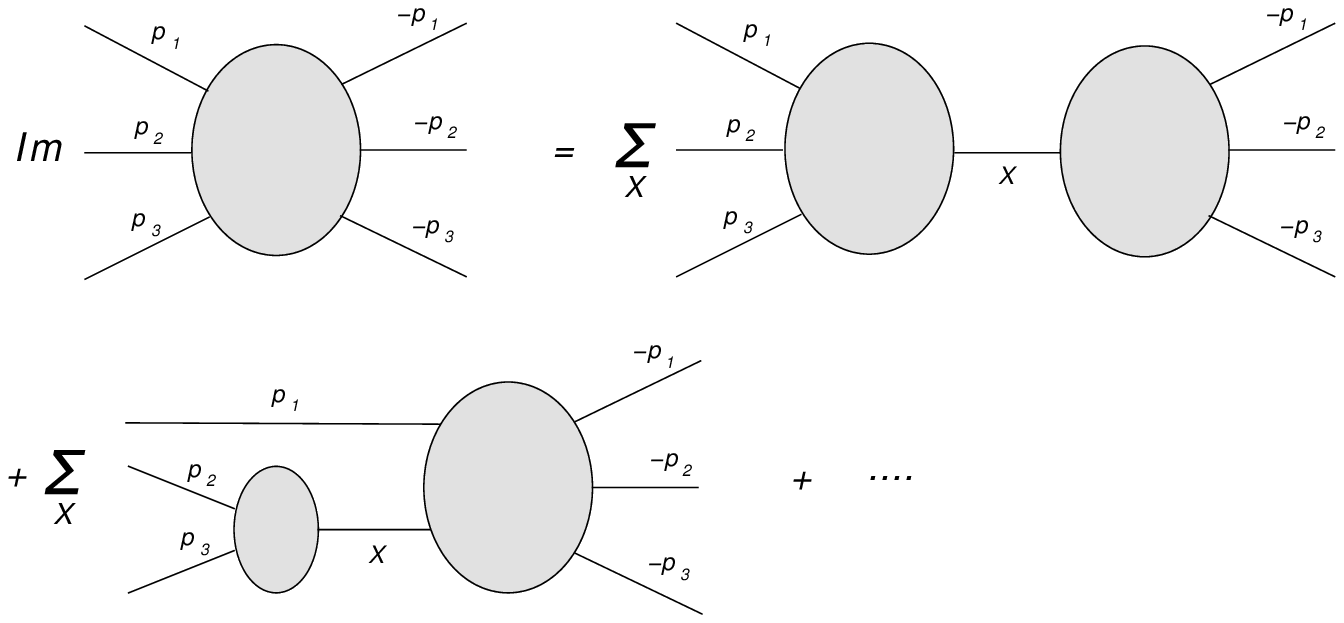}}

Proceeding by induction, we find that in unitary scale invariant theories
\eqn\matrixelements{\langle \Psi(p_1)\Psi(p_2)....\Psi(p_n)|Anything\rangle=0~.}
(Remember that all $p_i^2=0$.)

We thus learn that the on-shell dilaton is completely decoupled from unitary SFTs  in the sense that all the on-shell scattering amplitudes \matrixelements\ vanish.

\subsec{An Analogy with S-Matrix Theory}

Finally, we will use a familiar observation about S-matrix theory to conclude that such unitarity SFTs are actually CFTs. 

The S-matrix is invariant under field redefinitions. It is obviously trivial in free-field theory. In this subsection we will exploit the following assertion: if the S-matrix is trivial, then one can remove all the interactions in the theory by some change of variables. More precisely, we exploit the following claim: if the scattering of a certain one-particle state in a unitary theory is trivial, then after a suitable field redefinition this particle is described by a free field i.e.~there is a kinetic term in the Lagrangian but no interactions whatsoever.

Since the S-matrix is supposed to characterize the physical theory, 
this assertion sounds very plausible, but we do not know whether it can be established rigorously in general, and if so, under what precise assumptions (see especially the last example in section~4).
 
The dilaton coupling to the SFT is via $\int d^4x \Psi(x)T_\mu^\mu(x)+\cdots$.  As we explained in the introduction, we can imagine introducing a dilaton kinetic term $ f^2\int d^4x\half\Psi\square \Psi$ with a very large dimensionful coefficient $f$ (large compared to all the scales involved in the RG flow). If the coefficient of the kinetic term for $\Psi$ is very large, then the $n\rightarrow n$ scattering amplitudes of $\Psi$ quanta are given by $\CA_{2n}$, all of which have vanishing imaginary parts. Moreover, as our argument in the previous section has shown, due to unitarity, the transition amplitudes~\matrixelements\ vanish.
Hence, we expect that the interaction  $\int d^4x\, \Psi(x)T_\mu^\mu(x)+\cdots$ is removable by  a field redefinition of $\Psi$. This can be done if there exists a {\it local} operator $L$ such that $T_\mu^\mu=\square L$ ($L$ may vanish). In this case the unitary SFT is a CFT. 

Another way for the S-matrix to be trivial is that $T_\mu^\mu$ is a generalized free field of dimension $d$ (i.e. its $n$-point function is determined by Wick contractions)  and thus there are no connected diagrams with more than 2 insertions (the 2-point function is determined by dimensional analysis to be $p^4 \log p^2$).  Although it sounds like a  rather unlikely possibility in a local field theory, we have not been able to rule it out. 

{\it To summarize: unitary scale-invariant field theories must be either conformal field theories, or the trace of the energy-momentum tensor behaves like a generalized free field of dimension $d$.}

\newsec{Simple Examples}

In this section the ideas of this paper are illustrated in very simple examples. In all of the unitary theories discussed below, we indeed find that all the matrix elements~\matrixelements\ vanish, as predicted. However, in the free two-form theory with non-compact gauge symmetry, we will see that the vanishing of all the matrix elements~\matrixelements\ does not imply that the coupling to the dilaton can be removed by a change of variables (hence, this unitary theory is not conformal). We will comment on why this theory may be exceptional in this regard. By contrast, the free two-form theory with compact gauge symmetry is perfectly consistent with our claims.

\subsec{A Free Scalar Field}

Consider a free scalar field in four dimensions, $\varphi$. We can couple it to curved space in the minimal fashion
\eqn\freescalar{S=\half\int d^4x \sqrt g g^{\mu\nu}\del_\mu\varphi\del_\nu\varphi~.}
Let us specialize to metrics of the form $g_{\mu\nu}=(1+\Psi)^2\eta_{\mu\nu}$. Then the above action becomes 
\eqn\freescalari{S=\half\int d^4x \left( \del_\mu\varphi\del^\mu\varphi+2\Psi\del_\mu\varphi\del^\mu\varphi+\Psi^2 \del_\mu\varphi\del^\mu\varphi\right)~,}
where the indices are now contracted with the flat metric. It is  straightforward to compute the generating functional $W[\Psi]$ using conventional Feynman diagram techniques. We can further impose that $\square\Psi=0$, in which case the computation becomes identical to the scattering of  massless $\Psi$ particles weakly coupled to the original theory. Let us prove that for $\square\Psi=0$ the generating functional does not depend on $\Psi$, in accord with our general claims in the previous section. Note that this is somewhat nontrivial to see in terms of the Feynman diagrams generated by~\freescalari.

The leading interaction of $\Psi$ is with the operator $\del_\mu\varphi\del^\mu\varphi$, which is equivalent on-shell to $\half\square(\varphi^2)$. The difference between $\del_\mu\varphi\del^\mu\varphi$ and $\half\square(\varphi^2)$, being zero on-shell, can be absorbed by a redefinition of $\varphi$. This redefinition precisely cancels a  piece from the seagull term $\Psi^2(\del\varphi)^2$ such that $W[\Psi]$ is $\Psi$-independent for $\square \Psi=0$. In more detail, we rewrite~\freescalari\ as 
\eqn\freescalari{S=\half\int d^4x \left( \del_\mu\varphi\del^\mu\varphi+\square\Psi\varphi^2-2\Psi\varphi\square\varphi+\Psi^2 \del_\mu\varphi\del^\mu\varphi\right)~,}
which can be further rewritten as 
\eqn\freescalari{\eqalign{&S=\half\int d^4x \left( -\left(\varphi+\Psi\varphi\right)\square\left(\varphi+\Psi\varphi\right)+\square\Psi\varphi^2+\Psi\varphi\square(\Psi\varphi)+\Psi^2 \del_\mu\varphi\del^\mu\varphi\right)\cr&=\half\int d^4x \left( -\left(\varphi+\Psi\varphi\right)\square\left(\varphi+\Psi\varphi\right)+\square\Psi\varphi^2+\Psi\square\Psi\varphi^2\right)~,}}
where we used only integration by parts in the last line. Every interaction vertex with the source is now seen to contain an explicit $\square\Psi$. Hence, we can change the path integral variable to $\varphi'=\varphi+\Psi\varphi$ and we see that for $\square\Psi=0$ the amplitudes $\CA_n$ (and all the matrix elements~\matrixelements) vanish.
 
Furthermore, the coupling to the dilaton is through $\square \Psi\varphi^2$, which guarantees the theory has a traceless energy-momentum tensor after an improvement.
Of course, the fact that this theory is conformal is well known, the energy momentum tensor derived from~\freescalar\ is improvable and one can write the following symmetric, conserved, traceless EM tensor (in any $d>1$):
\eqn\freescalarii{T_{\mu\nu}=\del_\mu\varphi\del_\nu\varphi-\half \eta_{\mu\nu}\del^\rho\varphi\del_\rho\varphi+{2-d\over 4(d-1)}\left(\del_\mu\del_\nu-\eta_{\mu\nu}\del^2\right)\varphi^2~.}

\subsec{A non-Unitary Vector Field}

 Let us now present a simple example~\RivaGD\ which is scale invariant but it is not conformal invariant (see also~\ElShowkGZ). This example is non-unitary. 
Consider a massless vector field $A_\mu$  in $d$ dimensions (we do not impose gauge invariance). The most
general Lorenz-invariant quadratic action is
\eqn\cariva{S=\int d^d x\, (a\, \del_{\mu}A_{\nu} \del^{\mu}A^{\nu}+
b\, \del_{\mu}A_{\nu} \del^{\nu} A^{\mu})~.}
Of course, only the ratio of $a$ and $b$ matters for the results below.

It is easy to see that this theory always has a virial current
\eqn\virialcurrent{
V^\mu=(a+b)d\,A^\mu(\del^\nu A_\nu)-a(d-4)A_\nu F^{\mu\nu}~.
} 
To check whether it is a conformally invariant theory we have to look for the cases in which a local $L_{\mu\nu}$ can solve 
$V^\mu=\p_\nu L^{\nu\mu}$. One finds that a solution exists if and only if\foot{In fact, there is also a solution for $b=0$. One could simply write  (compare with \freescalarii)
$$T_{\mu\nu}\sim \del_{\mu}A_{\rho} \del_{\nu}A^{\rho}-\half\eta_{\mu\nu}\del_{\sigma}A_{\rho} \del^{\sigma}A^{\rho}+{2-d\over 4(d-1)} (\partial_\mu\partial_\nu-\eta_{\mu\nu}\partial^2) A^\rho A_\rho$$ which is symmetric, conserved, and traceless. However, it leads to twisted Lorentz charges, which act on $A^\mu$ as if these were $d$ Lorentz scalars. This is why the traceless energy-momentum tensor above should be disregarded. Thus, the special case $b=0$ corresponds to a scale invariant but {\it non-conformal} theory.} 
\eqn\comp{ b=-4a/d~.} 
Therefore, for this choice of coefficients, the theory becomes conformal. 
The explicit expression for $L_{\mu\nu}$ is 
\eqn\Lexpression{L^{\mu\nu}=(d-4)a\left(-A^\nu A^\mu+{1\over2}g^{\mu\nu}A^2\right)~.}
In the case~\comp\ $T_{\mu\nu}$ can be improved such that it is traceless.
Note that~\comp\ at $d=4$ leads to the usual Maxwell action which can be further rendered unitary by restricting to the gauge-invariant sector.

Let us remark that in the theories~\cariva\ our on-shell S-matrix elements $\CA_n$ should vanish identically for all $n$ in $d=4$, as we have proven on general grounds in subsection~3.1. But the transition amplitudes~\matrixelements\ may be nonzero because the underlying theory is non-unitary.\foot{It is easy to prove the vanishing of~$\CA_n$ by a trick that we borrow from~\LutyWW. We start from  $a=-b$, in which case the coupling to curved space is classically Weyl invariant (since the action coincides with that of the Maxwell theory). Thus for $a=-b$ the scattering of $\Psi$ quanta is clearly trivial. However, we can now study the theory in the $\xi$-gauges and obtain any of the other choices of $a,b$ in~\cariva. Since the new EM tensor only differs from the original one by $\delta_{BRST}$-exact terms, and since the ghosts (being improvable) are decoupled from $\Psi$, the scattering amplitudes $\CA_n$ are independent of $a,b$ and hence are trivial for any $a,b$ and for any $n$. Note that this does not imply anything about the complete set of matrix elements, since the gauge fixing potentially introduces $\xi$ dependence into the non-gauge invariant matrix elements.}

\subsec{A Free Scalar Field with Discrete Gauge Symmetry}

Here we discuss a compact scalar (one can think about it as a Nambu-Goldstone boson of a spontaneously broken $U(1)$). This is equivalent to a $(d-2)$-form gauge field with a compact gauge symmetry. The Lagrangian of a free scalar field with gauged shift symmetry is identical to the one of the usual non-compact free scalar field~\freescalar, only that $\varphi$ and $\varphi+2\pi f$ are gauge equivalent, where $f$ is some dimensionful scale.  

The theory is unitary, and the $\Psi$ scattering amplitudes and matrix elements are identical to those of  an ordinary non-compact free scalar field. These were explained to be trivial in subsection~4.1. 

However, naively the theory is not conformal in $d>2$ because the improvement term in~\freescalarii\ is not invariant under the gauge symmetry $\varphi\rightarrow \varphi+2\pi f$.  (In fact, for the same reason, the theory does not have a well-defined local scale current in $d>2$.\foot{Thus, the dual theory, a free gauge field in three dimensions has no local scale current. This has an interesting spinoff.  If there existed a local scale current, then by~\virial\ we could have immediately concluded that the $S^3$ partition function is independent of the radius of the $S^3$. However, an explicit computation reveals  a logarithmic dependence on the radius. If we now gap our free gauge field with a relevant Chern-Simons term with level $k$, then the logarithmic dependence on the radius above is replaced for a large sphere with a logarithm of $\sim e^2k$. This implies that the partition function must have a $\log k$, which is the familiar result for the partition function on $S^3$ of the Abelian Chern-Simons theory at level $k$~\refs{\SchwarzCN,\WittenHF}.})

Our argument that theories which are unitary and scale invariant must be conformal thus seems to be in tension with the apparent non-conformality of the Nambu-Goldstone boson.\foot{In $d=3$, a compact scalar is dual to a free Maxwell field with compact gauge symmetry. That the Maxwell field in $d=3$ is naively not conformal is already discussed, for example, in~\ElShowkGZ. Here we would like to emphasize that there is a crucial difference between the gauge field with compact and non-compact gauge symmetry. As we explain in the text, it is conformally invariant in the former case.} While our theorem that all the matrix elements~\matrixelements\ vanish is satisfied, naively, one cannot conclude that there is a local operator $L$ such that $T_\mu^\mu=\square L$.

The resolution is that $f$, which defines the radius of the Nambu-Goldstone model (and has a positive mass dimension for $d>2$), is an IR-irrelevant parameter -- it effectively goes to infinity at low energies. So, in the deep infrared, a Nambu-Goldstone is identical to the ordinary non-compact free scalar fields and, in particular, the operator $\varphi^2$ is a legal operator when we expand around some superselection sector. Of course, improving the energy momentum tensor by $\varphi^2$ around some superselection sector  breaks explicitly the global shift symmetry (which is already broken spontaneously by the choice of vacuum) but this is not forbidden.

There are several ways to see that  in $d>2$ the low energy limit of the compact scalar is the ordinary non-compact scalar. We could study the theory on $T^{d-1}\times R$. Then the zero mode on $T^{d-1}$ is a quantum mechanical degree of freedom with radius $\sim Vol(T^{d-1})f$. Localized wave functions on the circle look increasingly similar to those of the noncompact free scalar as we take the infrared limit. A more sophisticated check of this idea~\jh\ is to study the entanglement entropy of the compact scalar across a large $S^{d-2}$. One finds the familiar entanglement entropy of the ordinary non-compact massless scalar as the $S^{d-2}$ becomes very large. 

To conclude, the compact scalar theory is not a counter-example to our claims because in the infrared it does become conformal.

\subsec{A Free Scalar Field with Continuous Gauge Symmetry}

Let us now consider the following situation: a free scalar from which we remove the zero mode  by introducing a continuous gauge symmetry $\varphi\rightarrow\varphi+c$ for any real $c$. This is dual to a $(d-2)$-form gauge field with non-compact gauge symmetry. (In particular, in $d=3$ it is dual to the Maxwell field with non-compact gauge symmetry and in $d=4$ it is dual to the free two-form gauge field with non-compact gauge symmetry. See, for instance,~\ft)  Here again we find trivial matrix elements~\matrixelements, but since $\varphi^2$ is not gauge invariant, the theory is not conformal. Hence, even though all the matrix elements~\matrixelements\ are zero, the interaction with the dilaton cannot be removed by a change of variables.

We repeat: in this unitary theory the S-matrix for dilaton scattering is trivial, however, the interaction with the dilaton cannot be removed by a change of variables because the change of variables necessarily involves $\varphi^2$ which is not a well-defined local operator.

This may mean one of two things:

\medskip

\item{i.} Our main claim, that the triviality of the
scattering amplitudes of the $\Psi$-particle (3.6) implies there are new
variables such that the corresponding new field is free, is
generally correct. The precise formulation of this assertion may rely on some additional assumptions satisfied by generic ``good'' QFTs. 

\item{ii.} Our main claim, see above, is not necessarily satisfied by generic theories. 

\medskip

The second option is very implausible because, for sufficiently nice unitary QFTs, a trivial S-matrix {\it should} mean that the theory is free in some variables. Let us explain why the free two-form theory is likely a very special exception to this general rule, and therefore the first option holds true.

In flat space, the free two-form is closely related to the massless non-compact scalar. They have precisely the same Hilbert space. The difference is that to arrive at the free two-form we project out all the local operators in the free scalar theory that do not have derivatives acting on the scalar field. 
This projection out is of course the origin of the obstruction to having conformal invariance in the free two-form theory.
Since the Hilbert space is identical to the free non-compact scalar, and since the operators that are not projected out have precisely the same correlation functions as in the free noncompact scalar theory, the S-matrices for dilaton scattering are identical (i.e.~they are trivial in both cases).  

In general interacting QFT one should not be able to remove a subset of the local operators while retaining consistency, not modifying the Hilbert space, and not modifying the correlation functions of the operators that are not removed.  Even if the remaining set of operators is closed under the OPE, such a procedure generally leads to inconsistent models. For example, in two dimensions this leads to models that cease to have a sensible physical interpretation on various curved manifolds. 

Therefore, the counter-example of the free two-form theory can be understood as being related to the fact that in the case of a free massless scalar field in $d$ dimensions one can remove a subset of the local operators while not  jeopardizing the consistency of the theory. We do not expect this to be possible in general.

\centerline{\bf Acknowledgments}

We would like to thank N.~Arkani-Hamed, S.~Caron-Huot,  H.~Elvang, K.~Fredenhagen, D.~Freedman, D.~Jafferis, E.~Kiritsis, Y.~Nakayama, J.~Polchinski, S.~Pufu, R.~Rattazzi,  K.-H.~Rehren, N.~Seiberg, and S.~Yankielowicz for helpful comments and discussions.
AD would like to thank the Weizmann Institute of Science for
hospitality and gratefully acknowledges support from a Starting Grant of the European Research Council (ERC STG grant 279617).
ZK was supported by NSF grant PHY-0969448, a research grant from Peter and Patricia Gruber Awards,  a grant from Rosa and Emilio Segre research award, a grant from the Robert Rees Fund for Applied Research, and by the Israel Science Foundation under grant number~884/11. ZK  would also like to thank the United States-Israel Binational Science Foundation (BSF) for support under grant number~2010/629. This research was supported by the I-CORE Program of the Planning and Budgeting Committee and The Israel Science Foundation (grant NO 1937/12). AS would like to thank the Humboldt Foundation for support.
Any opinions, findings, and conclusions or recommendations expressed in this
material are those of the authors and do not necessarily reflect the views of the funding agencies.

\appendix{A}{A non-Diagonalizable Dilatation Operator}

Let us reconsider the current algebra equation~\currentzero\
\eqn\currentzeroappendix{i[\hat D, T_{\mu\nu}]=x^\rho\del_\rho T_{\mu
\nu}+d T_{\mu\nu}+\del^\rho\del^\sigma Y_{\rho\mu ; \sigma\nu}~.}
As discussed in section 2,  the matrix $\Gamma^I_J$ in~\currentii\ can be brought to a Jordan normal form. In the new basis a necessary condition for $Y_{\rho\mu ; \sigma\nu}$ in four dimensions to be non-removable from \currentzeroappendix\ is that its generalized eigenvalue is equal 2. 

Below we will show that in a unitary SFT $Y_{\rho\mu; \sigma\nu}$ from \currentzeroappendix\ of dimension 2 must be a scalar 
$Y_{\rho\mu; \sigma\nu}=(\eta_{\rho\sigma}\eta_{\mu\nu}-\eta_{\rho\nu}\eta_{\mu\sigma})Y$ or a conserved tensor, and trace of~\currentzeroappendix\ becomes 
\eqn\currentzeroappendixi{i[\hat D, T_{\mu\nu}]=x^\rho\del_\rho T_{\mu}^{
\mu}+d\, T_{\mu}^{\mu}+\square Y~.}
 
In the theories with non-canonical current algebra, i.e.~whenever $Y$ are present, we might not be able to use naive dimensional analysis to fix correlation functions (or their imaginary parts).  However, in case the current algebra is reduced to~\currentzeroappendixi\ the extra term does not affect the on-shell dilaton scattering amplitudes. Indeed the dilaton couples to the trace $T_{\mu}^{\mu}$, and the possible contrinution of the form $\square Y$  vanishes on-shell. Therefore, we can apply the dimensional analysis to the scattering amplitudes of $\Psi$ and the main argument of this paper holds i.e.~$T_\mu^\mu=\square L$ for some local $L$, while $L$ may mix with some $Y$ as the scale changes.

As a first step let us assume unitarity and re-derive the bound on the dimension of the  scalar operators in a SFT. We start with the most general  current algebra  
\eqn\currentzeroappendix{i[\hat D, \CO^I]=x^\rho\del_\rho \CO^I+\Gamma^I_J\CO^J~,}
and bring $\Gamma_J^I$ to a Jordan normal form. The diagonal elements which we call generalized dimensions will be denoted by $\Delta_i$. For each $\Delta_i$ there is at least one operator $\CO$ which sits at the bottom row of the $\Delta_i$ Jordan block  such that $i[\hat D, \CO]=x^\rho\del_\rho \CO+\Delta_i\CO$ i.e.~it is an ordinary eigenvector of the dilatation operator. 
Assuming vacuum is $\hat D$-invariant by dimensional analysis the two-point function of the scalars $\CO$ is fixed be\foot{In  case $\Delta_i-d/2$ is a non-negative integer  (B.4)  acquires an extra $\log p^2$.} \eqn\twopointm{\langle\CO(p)\CO(-p)\rangle={C p^{2(\Delta_i-d/2)}}~.}
The bound on the dimension $\Delta_i$ comes from re-writing \twopointm\ using K\"allen-Lehman representation 
\eqn\KL{{C p^{2(\Delta_i-d/2)}}=\int_0^\infty {d\mu^2 \rho(\mu)^2\over p^2-\mu^2+i\epsilon}~,}
and requiring that $\rho(\mu^2)$ is non-negative while the integral \KL\ converges~\refs{\CappelliYC,\GrinsteinQK}. The dimensional analysis implies $\rho(\mu^2)\sim \mu^{2(\Delta_i-d/2)}$ and hence $\Delta_i\ge (d-2)/2$ for \KL\ to converge. Of course one  can try to define the integral \KL\ via analytic continuation for $\rho(\mu^2)\sim \mu^{2\delta}$ and $\delta<-1$. The result rotated to the Euclidean space 
\eqn\twopoint{\langle\CO^i(x)\CO^i(0)\rangle={\tilde{C}\over x^{2\Delta_i}}}
would look sensible except for the fact that positive $\rho(\mu^2)$ would lead to a negative $\tilde{C}$ (for $-1>\delta>-2$). This violation of reflection positivity is a sign of a sickness  associated with  $\Delta_i$ that violates the unitarity bound.

A very similar line of reasoning works for non-scalar operators as well. The only important distinction is that the two-point function of say, vectors $\CO_\mu$, might have a term proportional to $p_\mu p_\nu p^{2(\Delta-d/2-1)}$. Convergence of \KL\ would now imply a stronger bound $\Delta\ge d/2$. In general a two-point function of spin $j$ operators might have the  term $p_{\mu_1} p_{\nu_1}\dots p_{\mu_j} p_{\nu_j}$ which would result in the bound $\Delta\ge d/2+j-1$ \GrinsteinQK. If the theory is conformal this term in fact must be present (and the real bound is even stronger). But if the theory is merely scale invariant not all of these terms might be present making the bound weaker. 
 For example the two-point function of traceless symmetric spin $2$ operators $\CO_{\mu\nu}$ could be made exclusively out of $\eta_{\mu\nu}$ and contain no ``$p_\mu p_\nu$'' terms:
\eqn\spintwo{\langle\CO_{\mu\nu}(p)\CO_{\mu'\nu'}(-p)\rangle\sim{\left(\eta_{\mu\nu}\eta_{\mu'\nu'}-{d\over2}(\eta_{\mu\mu'}\eta_{\nu\nu'}+\eta_{\mu\nu'}\eta_{\nu\mu'})\right) p^{2(\Delta-d/2)}}~.}
In such a case convergence of~\KL\ would imply the same bound as for the scalars $\Delta\ge (d-2)/2$. This is in fact too conservative, because the imaginary part of \spintwo\ is not sign definite and is not consistent with unitarity. In a unitary theory \spintwo\ must include the ``$p_\mu p_\nu$'' terms  which would strengthen the bound to be at least $\Delta \ge 2$ (and again this might be too conservative as well). Finally if dimension is exactly  $2$ the operator in question is conserved: the two-point function of both $\partial_\mu \partial_\n \CO_{\mu\nu}$  and $\square \CO_{\mu\nu}$ is a contact term implying these operators vanish. That is why $Y_{\mu\nu}$ can always be removed from \currentzeroappendixi.

\appendix{B}{More on Finite Counterterms}

In this appendix we discuss a  specialized class of renormalization group flows where more detailed information than what we have used in section~3 can be provided. Another motivation for this appendix is to present infinitely many new sum rules for the difference between the $a$-anomalies in RG flows connecting CFTs. We will emphasize an application of these infinitely many new sum rules for the problem of SFT vs CFT.

We consider renormalization group flows between two conformal field theories,  CFT$_{uv}$ and  CFT$_{ir}$. What we present in the following is a simple generalization of the ideas of~\KomargodskiVJ. We can couple the theory to a background metric in a diffeomorphism invariant fashion. We can then take the metric to be conformally flat $g_{\mu\nu}=e^{-2\tau}\eta_{\mu\nu}$. The effective action for the external dilaton source $\tau(x)$ is constrained by symmetries. The low energy effective action up to (and including) four derivatives is
\eqn\flatIR{\eqalign{S_{IR}=\int
d^4x\left(\half f^2e^{-2\tau}(\del\tau)^2+\kappa\left( \square \
\tau-(\del\tau)^2\right)^2
+(a_{uv}-a_{ir})\left(4(\del\tau)^2\square \
\tau-2(\del\tau)^4\right) \right)+\cdots~,}}
and the on-shell condition is $\square 
\tau-(\del\tau)^2=0$.
The term proportional to $\kappa$ does not contribute at the level of four derivatives. Therefore, any process of dilaton scattering is universally fixed at the level of four derivatives by the $a$-anomalies of the UV and IR CFTs. We can calculate the low energy limit of the $n-n$ scattering amplitude in the forward limit \eqn\scatteringamplitude{\CA_{2n}(s_{ij})=8\,{\tilde a_n\over f^{2 n}}\sum_{1\leq i<j\leq n}s_{ij}^2
\qquad\hbox{with}\qquad \tilde a_n={(2n-1)!\over 3!}\,(a_{uv}-a_{ir})~.}
Using these low energy scattering amplitudes we can write (convergent) sum rules.

The simplest sum rule is the one that corresponds to $n=2$
\eqn\sumrulesthree{a_{uv}-a_{ir}={f^4\over 4\pi}\int {ds\over s^3}\,\Im\CA_{4}(s)~.}
Here we see that it admits a natural generalization to any forward $n-n$ process  of dilaton scattering
\eqn\sumrulegenappendix{\tilde a_{n}\sum_{ij}s_{ij}^2={f^{2n}\over 4\pi}\int {d\lambda \over \lambda^3}\Im\CA_{2n}(\lambda s_{ij})~.}
This result is applicable for any RG flow between two CFTs.

Let us  explain how these sum rules can be applied for the problem of scale versus conformal invariance. Imagine that during the flow we can pass very close to an SFT. We imagine that there is a small parameter $\epsilon$ in the space of couplings such that we can get arbitrarily close to an SFT as we take $\epsilon$ to zero. See figure 3. We also imagine, for simplicity, that the infrared is gapped. The class of theories considered here  includes many interesting models, for example, any theory that can be reached by a deformation of the Gaussian fixed point (and subsequently deformed to an empty theory). 

\ifig\figten{We can probe potential SFTs by arranging a flow from a CFT that hovers near the SFT point for arbitrarily long RG time.} {\epsfxsize2.0in\epsfbox{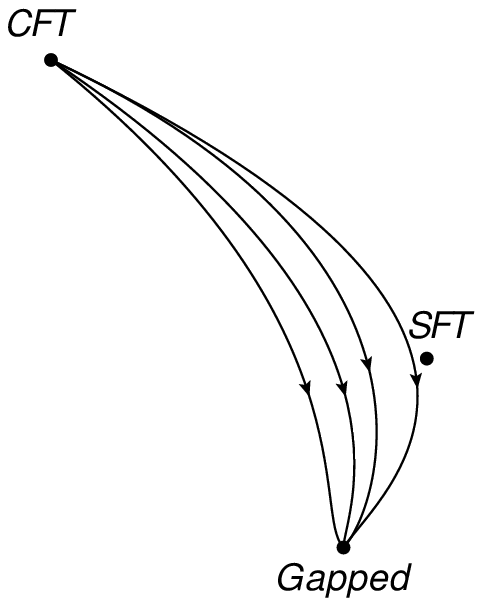}}

Having the small parameter $\epsilon$ that allows to hover near an SFT as in figure 3 means that in an energy range between $\mu_{IR}$ and $\mu_{UV}$, $\mu_{IR}\ll \mu_{UV}  $,  the theory is approximately scale invariant. Then,  by dimensional analysis, 
\eqn\ansatzSFT{\mu_{IR}^2\ll s_{ij}\ll \mu_{UV}^2, \ \mu_{IR}^2\ll\lambda s_{ij}\ll \mu_{UV}^2:\qquad \Im\CA_n(\lambda s_{ij})= \lambda^2 \CF_n(s_{ij})~.}
If the function $\CF_n(s_{ij})$ is non-vanishing, such a behavior leads to a contradiction with the sum rules~\sumrulegenappendix\ because the sum rules cease to converge as $\epsilon\rightarrow0$. Therefore, the imaginary part~\ansatzSFT\ vanishes for all $n$. 
For example, for $\CA_4(s)$ we might have expected by dimensional analysis that $\CA_4(s)=\kappa s^2\log\left({s\over \mu_{IR}^2}\right)$ in the SFT regime, but the coefficient must necessarily vanish $\kappa=0$.

Repeating the argument of section 3 we find that, imposing unitarity, all the matrix elements~\matrixelements\ vanish in the SFT energy range between $\mu_{IR}$ and $\mu_{UV}$.
The coupling of the dilaton to the SFT in this energy range can be consistent with the vanishing of all the connected matrix elements if the theory is a CFT (in which case the dilaton decouples on-shell), or if the trace of the energy-momentum tensor is a generalized free field (in which case there are no connected diagrams).

\listrefs
\end